\documentclass[12pt,a4paper]{article}
\usepackage{axodraw}
\makeatletter
\if@twoside \oddsidemargin -17pt \evensidemargin 00pt
\else \oddsidemargin 00pt \evensidemargin 00pt
\fi
\expandafter\ifx\csname mathrm\endcsname\relax\def\mathrm#1{{\rm #1}}\fi

\newcommand{\ps}{p\hspace{-0.42em}/}
\newcommand{\rs}{r\hspace{-0.42em}/}
\newcommand{\qs}{q\hspace{-0.5em}/}

\renewcommand{\theequation}{\thesection.\arabic{equation}}
\newcounter{saveeqn}

\@addtoreset{equation}{section}

\makeatother

\def\nl{\nonumber\\}

\def\nln{\nonumber\\*[-1ex]\phantom{\fbox{\rule{0em}{2ex}}}}

\def\beq{\begin{equation}}
\def\eeq{\end{equation}}
\def\beqar{\begin{eqnarray}}
\def\eeqar{\end{eqnarray}}
\def\barr#1{\begin{array}{#1}}
\def\earr{\end{array}}
\def\bfi{\begin{figure}}
\def\efi{\end{figure}}
\def\btab{\begin{table}}
\def\etab{\end{table}}
\def\bce{\begin{center}}
\def\ece{\end{center}}

\def\text{\textstyle}

\def\de{\delta}
\def\veps{\varepsilon}
\def\la{\lambda}
\def\si{\sigma}
\def\Ga{\Gamma}
\def\De{\Delta}

\def\refeq#1{\mbox{(\ref{#1})}}

\def\refse#1{\mbox{Sect.~\ref{#1}}}
\def\refses#1{\mbox{Sects.~\ref{#1}}}

\def\refapp#1{\mbox{App.~\ref{#1}}}
\def\citere#1{\mbox{Ref.~\cite{#1}}}
\def\citeres#1{\mbox{Refs.~\cite{#1}}}

\def\solid{\raise.9mm\hbox{\protect\rule{1.1cm}{.2mm}}}
\def\dash{\raise.9mm\hbox{\protect\rule{2mm}{.2mm}}\hspace*{1mm}}

\newcommand{\TeV}{\unskip\,\mathrm{TeV}}

\def\mathswitchr#1{\relax\ifmmode{\mathrm{#1}}\else$\mathrm{#1}$\fi}

\newcommand{\PW}{\mathswitchr W}
\newcommand{\PZ}{\mathswitchr Z}
\newcommand{\PA}{\mathswitchr A}

\newcommand{\PH}{\mathswitchr H}

\newcommand{\Pb}{\mathswitchr b}

\newcommand{\Pt}{\mathswitchr t}

\newcommand{\PWp}{\mathswitchr {W^+}}
\newcommand{\PWm}{\mathswitchr {W^-}}

\def\mathswitch#1{\relax\ifmmode#1\else$#1$\fi}

\newcommand{\MW}{\mathswitch {M_\PW}}

\newcommand{\MZ}{\mathswitch {M_\PZ}}
\newcommand{\MH}{\mathswitch {M_\PH}}

\newcommand{\Mt}{\mathswitch {m_\Pt}}

\newcommand{\scrs}{\scriptscriptstyle}
\newcommand{\sw}{\mathswitch {s_{\scrs\PW}}}

\newcommand{\bew}{b^{\ew}}

\newcommand{\cew}{C^{\ew}}

\newcommand{\ckm}{{\bf V}}

\def\ie{i.e.\ }

\def\cf{cf.\ }

\renewcommand{\O}{{\cal O}}

\newcommand{\LA}{\stackrel{\mathrm{LA}}{=}}

\newcommand{\SUtwo}{\mathrm{SU(2)}}
\newcommand{\Uone}{\mathrm{U}(1)}
\newcommand{\ewgroup}{\SUtwo\times\Uone}

\newcommand{\rR}{\mathrm{R}}
\newcommand{\rL}{\mathrm{L}}
\newcommand{\rT}{{\mathrm{T}}}

\newcommand{\ri}{\mathrm{i}}

\newcommand{\coll}{{\mathrm{coll}}}

\newcommand{\ew}{{\mathrm{ew}}}

\newcommand{\M}{{\cal {M}}}
\renewcommand{\L}{{\cal L}}

\newcommand{\cc}{{\mathrm{C}}}

\newcommand{\ddl}{\frac{\mathrm{d}^Dl}{(2\pi)^D}}
\newcommand{\ddq}{\frac{\mathrm{d}^Dq}{(2\pi)^D}}
\newcommand{\brs}{{\it s}}

\newcommand{\Psibar}{\bar{\Psi}}

\makeatletter
\newcount\@tempcntc
\def\@citex[#1]#2{\if@filesw\immediate\write\@auxout{\string\citation{#2}}\fi
  \@tempcnta\z@\@tempcntb\m@ne\def\@citea{}\@cite{\@for\@citeb:=#2\do
    {\@ifundefined
       {b@\@citeb}{\@citeo\@tempcntb\m@ne\@citea
        \def\@citea{,\penalty\@m\ }{\bf ?}\@warning
       {Citation `\@citeb' on page \thepage \space undefined}}%
    {\setbox\z@\hbox{\global\@tempcntc0\csname
b@\@citeb\endcsname\relax}%
     \ifnum\@tempcntc=\z@ \@citeo\@tempcntb\m@ne
       \@citea\def\@citea{,\penalty\@m}
       \hbox{\csname b@\@citeb\endcsname}%
     \else
      \advance\@tempcntb\@ne
      \ifnum\@tempcntb=\@tempcntc
      \else\advance\@tempcntb\m@ne\@citeo
      \@tempcnta\@tempcntc\@tempcntb\@tempcntc\fi\fi}}\@citeo}{#1}}

\def\@citeo{\ifnum\@tempcnta>\@tempcntb\else\@citea
  \def\@citea{,\penalty\@m}%
  \ifnum\@tempcnta=\@tempcntb\the\@tempcnta\else
   {\advance\@tempcnta\@ne\ifnum\@tempcnta=\@tempcntb \else
\def\@citea{--}\fi
    \advance\@tempcnta\m@ne\the\@tempcnta\@citea\the\@tempcntb}\fi\fi}
\makeatother

\newcommand{\vev}{{\bf v}}

\def\draftdate{\relax}
\def\mda{\relax}
\def\mua{\relax}
\def\mla{\relax}
\def\draft{
\def\thtystars{******************************}
\def\sixtystars{\thtystars\thtystars}
\typeout{}
\typeout{\sixtystars**}
\typeout{* Draft mode!
         For final version remove \protect\draft\space in source file *}
\typeout{\sixtystars**}
\typeout{}
\def\draftdate{\today}
\def\mua{\marginpar[\boldmath\hfil$\uparrow$]%
                   {\boldmath$\uparrow$\hfil}%
                    \typeout{marginpar: $\uparrow$}\ignorespaces}
\def\mda{\marginpar[\boldmath\hfil$\downarrow$]%
                   {\boldmath$\downarrow$\hfil}%
                    \typeout{marginpar: $\downarrow$}\ignorespaces}
\def\mla{\marginpar[\boldmath\hfil$\rightarrow$]%
                   {\boldmath$\leftarrow $\hfil}%
                    \typeout{marginpar: $\leftrightarrow$}\ignorespaces}
\def\Mua{\marginpar[\boldmath\hfil$\Uparrow$]%
                   {\boldmath$\Uparrow$\hfil}%
                    \typeout{marginpar: $\Uparrow$}\ignorespaces}
\def\Mda{\marginpar[\boldmath\hfil$\Downarrow$]%
                   {\boldmath$\Downarrow$\hfil}%
                    \typeout{marginpar: $\Downarrow$}\ignorespaces}
\def\Mla{\marginpar[\boldmath\hfil$\Rightarrow$]%
                   {\boldmath$\Leftarrow $\hfil}%
                    \typeout{marginpar: $\Leftrightarrow$}\ignorespaces}
\overfullrule 5pt
\oddsidemargin -15mm
\marginparwidth 29mm
}

\makeatletter

\def\eqnarray{\stepcounter{equation}\let\@currentlabel=\theequation
\global\@eqnswtrue
\global\@eqcnt\z@\tabskip\@centering\let\\=\@eqncr
$$\halign to \displaywidth\bgroup\hskip\@centering
  $\displaystyle\tabskip\z@{##}$\@eqnsel&\global\@eqcnt\@ne
  \hskip 2\arraycolsep \hfil${##}$\hfil
  &\global\@eqcnt\tw@ \hskip 2\arraycolsep $\displaystyle\tabskip\z@{##}$\hfil
   \tabskip\@centering&\llap{##}\tabskip\z@\cr}
\def\appendix{\par
 \setcounter{section}{0} \setcounter{subsection}{0}
 \def\thesection{\Alph{section}}}

\makeatother

\begin{document}

\thispagestyle{empty}
\def\thefootnote{\fnsymbol{footnote}}
\setcounter{footnote}{1}
\null
\draftdate\hfill  PSI-PR-01-06 
\\
\strut\hfill ZU-TH-13-01\\
\strut\hfill hep-ph/0104127
\vskip 0cm
\vfill
\begin{center}
{\Large \bf
One-loop leading logarithms\\ in electroweak radiative corrections\\ 
II. Factorization of collinear singularities
\par} \vskip 2.5em
{\large
{\sc A.~Denner%
}\\[1ex]
{\normalsize \it Paul Scherrer Institut\\
CH-5232 Villigen PSI, Switzerland}\\[2ex]
{\sc S.~Pozzorini%
}\\[1ex]
{\normalsize \it
Institute of Theoretical Physics\\ University of Z\"urich, Switzerland \\[2ex]and \\[2ex]
{\normalsize \it Paul Scherrer Institut\\
CH-5232 Villigen PSI, Switzerland}\\[2ex]
}}

\par \vskip 1em
\end{center}
\par
\vskip .0cm 
\vfill {\bf Abstract:} \par 

We discuss the evaluation of the collinear single-logarithmic
contributions to virtual electroweak corrections at high energies.
More precisely, we proof the factorization of the mass singularities
originating from loop diagrams involving collinear virtual gauge
bosons coupled to external legs.  We discuss, in particular, processes
involving external longitudinal gauge bosons, which are treated using
the Goldstone-boson equivalence theorem.  The proof of factorization
is performed within the 't~Hooft--Feynman gauge at one-loop order and
applies to arbitrary electroweak processes that are not
mass-suppressed at high energies.  As basic ingredient we use Ward
identities for Green functions with arbitrary external particles
involving a gauge boson collinear to one of these. The Ward identities
are derived from the BRS invariance of the spontaneously broken
electroweak gauge theory.

\par
\vskip 1cm
\noindent
April 2001 
\par
\null
\setcounter{page}{0}
\clearpage
\def\thefootnote{\arabic{footnote}}
\setcounter{footnote}{0}

\section{Introduction}
In the energy range above the electroweak scale, $\sqrt{s}\gg \MW$,
electroweak radiative corrections are dominated by double-logarithmic
(DL) terms of the form $\alpha\log^2{(s/\MW^2)}$ and
single-logarithmic (SL) terms of the form $\alpha\log{(s/\MW^2)}$
involving the ratio of the energy to the electroweak scale (see
\citeres{Ciafaloni:2000rp,Kuhn:2000nn,Fadin:2000bq,Melles:2001gw,Beenakker:2000kb,Hori:2000tm,Denner:2001jv,Beccaria:2000fk}
and references therein).  Such corrections grow with energy, and at
$\sqrt{s}=0.5$--$1\TeV$ they are typically of order $10\%$ of the
theoretical prediction.  In the TeV range, the SL terms are
numerically of the same size as the DL terms.

For electroweak processes that are not mass-suppressed at high
energies, these leading logarithmic corrections are universal.  On the
one hand, single logarithms originating from short-distance scales
result from the renormalization of dimensionless parameters, \ie the
running of the gauge, Yukawa, and scalar couplings. On the other hand,
universal logarithms originating from the long-distance scale
$\MW\ll\sqrt{s}$ are expected to factorize, \ie they can be associated
with external lines or pairs of external lines in Feynman diagrams.  
They consist of DL and SL terms
originating from soft-collinear and collinear (or soft) gauge bosons,
respectively, coupled to external legs. The non-logarithmic terms are
in general
non-universal and have to be evaluated for each process separately if
needed.

In the recent literature (see \citeres{Ciafaloni:2000rp,Kuhn:2000nn,Fadin:2000bq,Melles:2001gw,Beenakker:2000kb,Hori:2000tm,Denner:2001jv,Beccaria:2000fk} and references therein),
most interest has been devoted to electroweak long-distance
corrections, which have often been compared to the well-known soft and
collinear singularities observed in QCD (see for instance
\citere{Collins:1989gx}).  This is a useful guide-line in order to
understand universal effects, and also to discuss specific features
that distinguish a spontaneously broken gauge theory from a symmetric
one.

The main difference between QCD and the Electroweak Standard Model is that
the masses of the weak gauge bosons provide a physical cut-off for
real $\PZ$- and $\PW$-boson emission.  Therefore, for a sufficiently
good experimental resolution, soft and collinear weak-boson radiation
need not be included in the theoretical predictions 
and, except for electromagnetic real corrections,
we can restrict ourselves to large logarithms originating from virtual
corrections.
 
Here we concentrate on the factorization of virtual
collinear corrections in high-energy electroweak reactions.  In QCD,
factorization is strictly connected to gauge symmetry
\cite{Collins:1989gx}.  Therefore, it is natural to ask if and how
factorization is affected by the spontaneous breaking of the gauge
symmetry within the electroweak theory.

In the literature \cite{Melles:2000ia}, this question has been avoided
by assuming that ``the electroweak theory is in the symmetric phase at
high energies''.  In this case, one restricts oneself to the symmetric
part of the electroweak Lagrangian ($\mathcal{L_{\mathrm{symm}}}$),
which corresponds to a vanishing vacuum expectation value (vev) of the
scalar doublet and depends only on dimensionless parameters;
gauge-boson masses in the propagators act merely as infrared cut-off.
In this ``symmetric approach'', methods and results obtained within
QCD are extended to the electroweak theory
\cite{Fadin:2000bq,Melles:2001gw,Melles:2000ia}.  Under these
assumptions,
only the following specifically electroweak ingredients need to be
included:
\begin{itemize}
\item Yukawa and scalar sector:  
since the dimensionless
  Yukawa and scalar couplings are proportional to the fermion and
  Higgs-boson masses, respectively, their effects are enhanced if these
  particles are heavy.  Especially, one finds large logarithmic
  corrections proportional to $\Mt^2/\MW^2\log{(s/\MW^2)}$ for
  processes involving 
heavy quarks or Higgs bosons.
\item Mixing of neutral gauge bosons: the neutral mass-eigenstate
  gauge bosons $\PA$ and $\PZ$ originate from mixing between the
  $\Uone$ and $\SUtwo$ eigenstates.  Since the adjoint representation
  of the $\ewgroup$ group is not irreducible, factorization is
  non-diagonal for processes involving external photons and
  $\PZ$~bosons. Note that the definition of the mass eigenstates requires to
  consider the theory at the electroweak scale.
\end{itemize}
The symmetric approach seems to be adequate for  electroweak
processes involving only fermions, transverse gauge bosons, and
Higgs bosons as external particles,
since these states are 
already present in the symmetric phase.
However, it is less clear whether this approach is adequate for
processes involving longitudinal gauge bosons, which
originate from spontaneous symmetry breaking.

For a rigorous treatment and, in particular, for processes involving
arbitrary external fields corresponding to mass eigenstates of the
electroweak theory, we need a ``complete electroweak'' approach.
Therefore, we calculate the leading logarithmic one-loop corrections
that originate from the complete Lagrangian, including terms
proportional to the vev.  However, we restrict ourselves to processes
that are not mass-suppressed in the high-energy limit, \ie processes
originating from $\mathcal{L_{\mathrm{symm}}}$ in lowest-order.  A
process is called mass-suppressed if its matrix element with mass
dimension $d$ does not scale as $E^d$ in the high-energy limit $E\gg
\MW$ 
but with $E^{d-n}\MW^{n}$, $n>0$. To proof the
factorization of the virtual collinear single logarithms, we use Ward
identities that are based on the symmetry of the complete Lagrangian.

In particular, we discuss the effects that are related to the part of
the Lagrangian that results from spontaneous symmetry breaking
($\L_{v}$), \ie the part proportional to the non-vanishing vev.  The
part $\L_{v}$ consists of terms that
are bilinear and trilinear in
the fields. In lowest order, bilinear terms in the scalar sector provide
gauge-boson masses and mixing between gauge bosons and would-be Goldstone
bosons.  Corresponding mixing terms are introduced in the 't~Hooft
gauge-fixing Lagrangian.  As a consequence of the BRS invariance, the
mixing terms lead to the well-known Goldstone-boson equivalence
theorem (GBET) \cite{et}, which relates longitudinal gauge bosons to
would-be Goldstone bosons in the high-energy limit.  Beyond tree-level,
also the trilinear couplings with mass dimension in $\L_{v}$ have to
be taken into account, since they give leading SL corrections to the
mass- and mixing-terms, and thus corrections to the GBET (for the
corrections to the GBET see \citere{etcorr}).

The complete one-loop results for high-energy leading electroweak DL
and SL correcti\-ons have been presented in \citere{Denner:2001jv}.
They include soft-collinear, purely collinear, purely soft, as well as
parameter-renormalization contributions.  In this article we
concentrate on the purely collinear SL corrections. Especially, we
proof the non-trivial factorization of the part originating from
mass-singular loop diagrams in the 't~Hooft--Feynman gauge.  In
\refse{masssing} (and \refapp{app:masssing}) we discuss mass
singularities originating from loop diagrams and show that they are
restricted to virtual gauge bosons coupled to external lines.  The
factorization of these mass singularities is demonstrated
in \refse{sec:Fact} using {\em collinear Ward identities}. We also
recall the complete gauge-invariant results for the collinear and soft
single logarithmic corrections given in \citere{Denner:2001jv},
including the part originating from renormalization
(field-renormalization constants and corrections to the GBET).  The
collinear Ward identities, which constitute the basis for the proof,
are derived in \refse{CWIsection} using the BRS invariance of the
electroweak theory (\refapp{BRStra}). Our conventions for Green
functions can be found in \refapp{app:GFs}.

\section{Collinear mass singularities}
\label{masssing}

\subsection{Notation}

\newcommand{\Oper}{O}
We consider electroweak processes involving $n$ arbitrary external
particles.  Lowest-order (LO)
matrix elements 
are denoted by
\begin{equation}\label{Bornampli}
\M_0^{\varphi_{i_1} \ldots \varphi_{i_n}}(p_1,\ldots, p_n),
\end{equation}
where all momenta are considered to be incoming.  The (incoming) 
fields
$\varphi_{i_k}$ represent physical fields in the standard model, \ie
fields corresponding to mass eigenstates for fermions, gauge bosons,
or Higgs bosons.  Longitudinal gauge bosons are replaced by the
corresponding would-be Goldstone bosons via the Goldstone-boson
equivalence theorem (GBET).
In the limit where all external momenta $p_k$ are on-shell, and all other 
invariants are much larger than the gauge-boson masses, \ie 
\beq \label{Sudaklim} 
\left(\sum_{l=1}^N p_{k_l}\right)^2\sim s \gg\MW^2, \qquad 1<N<n-1,\qquad
{k_l}\ne {k_{l'}} \ \mbox{ for }\ l\ne l',
\eeq
the one-loop corrections to \refeq{Bornampli} receive large
mass-singular logarithmic contributions.  Here, we assume that all
invariants are of the order $s$, the square of the typical energy
scale of the considered process, and we restrict ourselves to purely
collinear contributions containing terms of the form 
$\alpha\log{(s/M^2)}$, where $M$ is equal to $\MW$ or to a
light-fermion mass.  We show that these corrections $\de^{\cc}
\M^{\varphi_{i_1} \ldots \varphi_{i_n}}$ 
factorize and can be associated to the external states,
\beq\label{subllogfact}
\de^{\cc} \M^{\varphi_{i_1} \ldots \varphi_{i_n}} =\sum_{k=1}^n \sum_{\varphi_{i'_k}}\delta^\cc_{\varphi_{i'_k}\varphi_{i_k}}
\M_0^{\varphi_{i_1} \ldots \varphi_{i'_k} \ldots \varphi_{i_n}}.
\eeq
This universal (process-independent) result has been obtained within
the 't~Hooft--Feyn\-man gauge, using the independence of the $S$ matrix
of the scale $\mu$ of dimensional regularization \cite{Denner:2001jv}.
For external fermions, transverse gauge bosons, and Higgs bosons, the
large logarithms are isolated in
the $\mu$-dependent part of field
renormalization constants (FRC's) $\de Z$ and universal collinear
factors $\delta^\coll$ from mass-singular loop diagrams,
\beq\label{subllogfact2}
\delta^\cc_{\varphi_{i'_k}\varphi_{i_k}}=\left.\left(\frac{1}{2}\delta
  Z_{\varphi_{i'_k}
    \varphi_{i_k}}+\delta^\coll_{\varphi_{i'_k}\varphi_{i_k}}\right)\right|_{\mu^2=s}.
\eeq
External longitudinal gauge bosons $V_\rL^b=Z_\rL,W^\pm_\rL$ are
related to the corresponding would-be Goldstone bosons
$\Phi_b=\chi,\phi^\pm$ using the GBET.  The corresponding
collinear corrections are given by
\beq\label{subllogfact3}
 \delta^\cc_{V_\rL^{b'}V_\rL^{b}}=\left.\left(\de_{V^{b'}V^{b }}\de
   C_{\Phi_{b}}
+ \delta^\coll_{\Phi_{b'}\Phi_{b}}\right)\right|_{\mu^2=s},
\eeq
and depend on the collinear factors for would-be Goldstone bosons
and on the corrections $\de C_{\Phi_{b}}$ 
to the GBET.  These
latter contain the FRC's for gauge bosons, longitudinal self-energy
and mixing-energy contributions, and mass counterterms
\cite{Denner:2001jv}.

The FRC's and the corrections to the GBET factorize in an obvious way.
Explicit results for these contributions have been presented in
\citere{Denner:2001jv}. In the following, we discuss only the
non-trivial factorization of mass-singular truncated 
loop diagrams leading
to the collinear factors $\de^\coll$.

\subsection{Mass singularities in loop diagrams}
As has been proved by Kinoshita \cite{Kinoshita:1962ur}, mass-singular
logarithmic corrections arise from loop diagrams where an external
on-shell line splits into two collinear 
internal lines,
\beqar\label{colldiagram}
\vcenter{\hbox{\begin{picture}(85,50)(0,-25)
\Line(5,0)(25,0)
\Line(25,0)(60.9,14.1)
\Line(25,0)(60.9,-14.1)
\Vertex(25,0){2}
\GCirc(65,0){15}{1}
\Text(40,-15)[t]{\scriptsize $\varphi_j$}
\Text(5,5)[lb]{\scriptsize $\varphi_{i}$}
\Text(40,15)[b]{\scriptsize $\varphi_{k}$}
\end{picture}}}.
\eeqar
Here and in the following, all on-shell external lines that are not
involved in our argumentation are omitted 
in the graphical
representation.
The diagrams have to be understood as truncated; the self-energy
insertions in external legs and the corresponding mass singularities 
enter the FRC's in \refeq{subllogfact2}.

We
consider splittings $\varphi_i(p)\rightarrow
\varphi_j(q)\varphi_k(p-q)$ involving arbitrary combinations of
fields. These lead to loop integrals of the type
\newcommand{\ddqt}{\frac{\mathrm{d}^{D-2}q_\rT}{(2\pi)^{D-2}}}
\newcommand{\ddqtz}{\frac{\mathrm{d}^{2-2\varepsilon}q_\rT}{(2\pi)^{2-2\varepsilon}}}
\beqar \label{masssingloop2}
I&=&
-\ri (4\pi)^2\mu^{4-D} \int\ddq
\frac{N_{ijk}(q)}{(q^2-M_{j}^2+\ri \varepsilon)[(p-q)^2-M_{k}^2+\ri \varepsilon]}.  
\eeqar
The part denoted by $N_{ijk}(q)$ is kept implicit. It
consists of the LO contribution from the ``white blob'' in
\refeq{colldiagram}, of the wave-function 
(spinor or
polarization vector) corresponding to the external line $\varphi_i$,
of the $\varphi_i\varphi_j\varphi_k$ vertex, and of the numerators of
the $\varphi_j$ and $\varphi_k$~propagators.  
Since the soft contributions can be treated in the eikonal approximation
\cite{Denner:2001jv}, we assume that the part of $N_{ijk}(q)$ that is
singular in the soft limits $q^\mu\rightarrow 0$ and $q^\mu\rightarrow
p^\mu$ has been subtracted
(see \refse{sec:Fact}).

The mass singularity in \refeq{masssingloop2} originates from the
denominators of the $\varphi_j$ and $\varphi_k$~propagators in the
collinear region $q^\mu\rightarrow xp^\mu$. This is discussed in
detail in \refapp{app:masssing}, where we show that the mass
singularity can be extracted from \refeq{masssingloop2} by treating
the integrand $N_{ijk}(q)$ in the collinear approximation
\refeq{collappdef}.  The resulting contribution reads
\beqar \label{masssingloop2c}
I&\LA&
\log{\left(\frac{\mu^2}{M^2}\right)}\int_0^1\mathrm{d}x \, 
N_{ijk}(xp)
\eeqar
in logarithmic approximation (LA),
where $M^2\sim\max{(p^2,M_j^2,M_k^2)}$.
Since we consider all masses $\MW$, $\MZ$, $\MH$, and $\Mt$ to be of
the same order of magnitude, the scale
$M$ is either given by $\MW$ or by a light-fermion mass.

If we now apply the collinear approximation \refeq{collappdef}
to all splittings
$\varphi_i\rightarrow \varphi_j\varphi_k$, which are allowed by the
electroweak Feynman rules
\cite{DennBohmJos},
it turns out that $N_{ijk}$ is
mass-suppressed in most of the cases.  
This can be easily verified by looking at the external
part of the diagram \refeq{colldiagram}, containing the 
$\varphi_i$ wave function,
the $\varphi_i\varphi_j\varphi_k$ vertex, and the numerators of the
$\varphi_j$ and $\varphi_k$ propagators.  
Many contributions are proportional to $M$, $p^2$, $p_\mu\veps^\mu(p)$,
 or $\ps u(p)$ and thus mass-suppressed.
Consider as an example the
case $V_\rT\rightarrow\Psi\bar{\Psi}$, where a transverse gauge boson
splits into a fermion--antifermion pair. Here
\beq
N(q)\propto  
\varepsilon_\rT^\mu(p)(\ps-\qs)\gamma_\mu \qs \longrightarrow
x(1-x)\varepsilon_\rT^\mu(p)(2p_\mu\ps-p^2\gamma_\mu)
\eeq
is mass-suppressed in the collinear limit, $q^\mu\rightarrow xp^\mu$,
owing to $p_\mu\varepsilon_\rT^\mu(p)=0$ and $p^2\ll s$.  Similar
suppressions occur in all cases, except for the splittings
$\varphi_{i} \rightarrow V^a \varphi_{i'}$ where a virtual gauge boson
$V^a=A,Z,W^\pm$ is emitted and $\varphi_{i}$ and $\varphi_{i'}$
are both fermions, gauge bosons, or scalars.
These unsuppressed splittings are considered in \refse{sec:Fact}.


\section{Factorization of collinear singularities}
\label{sec:Fact}
In this section, we evaluate the loop diagrams \refeq{colldiagram}
involving  splittings 
\beq
\varphi_{i_k}(p_k) \rightarrow V_\mu^a(q) \varphi_{i'_k}(p_k-q).
\eeq
As mentioned in the previous section, we subtract soft contributions
that give rise to singularities of the integrand $N(q)$ in the region
$q^\mu\rightarrow 0$.  These result from diagrams where the gauge boson
$V^a$ couples to another external leg $\varphi_{i_l}$ in the eikonal
approximation (the second term in the second line of
\refeq{collfactorization}). 
These soft contributions can be treated separately
\cite{Denner:2001jv}.  For the remaining SL collinear singularities we
derive the factorization identities
\beqar\label{collfactorization}
\lefteqn{
\delta^\coll \M^{\varphi_{i_k}}(p_k)=}\qquad\nl
&=&\sum_{V^a=A,Z,W^\pm} \sum_{\varphi_{i'_k}}\left\{
\left[
\vcenter{\hbox{\begin{picture}(85,60)(0,-30)
\Line(5,0)(25,0)
\Line(25,0)(60.9,14.1)
\Photon(25,0)(60.9,-14.1){-2}{3}
\Vertex(25,0){2}
\GCirc(65,0){15}{1}
\Text(35,-14)[lt]{\scriptsize$V^a$}
\Text(35,+12)[lb]{\scriptsize$\varphi_{i'_k}$}
\Text(5,5)[lb]{\scriptsize $\varphi_{i_k}$}
\end{picture}}}
\right]_{\mathrm{trunc.}}
- \sum_{l\neq k} \sum_{\varphi_{i'_l}}\left[
\vcenter{\hbox{\begin{picture}(80,80)(5,-40)
\Line(65,0)(25,30)
\Line(65,0)(25,-30)
\PhotonArc(65,0)(32.5,143.13,216.87){2}{4.5}
\Vertex(39,19.5){2}
\Vertex(39,-19.5){2}
\GCirc(65,0){15}{1}
\Text(30,0)[r]{\scriptsize$V^a$}
\Text(20,27)[rb]{\scriptsize$\varphi_{i_k}$}
\Text(20,-27)[rt]{\scriptsize$\varphi_{i_l}$}
\Text(46,19)[lb]{\scriptsize$\varphi_{i'_k}$}
\Text(46,-19)[lt]{\scriptsize$\varphi_{i'_l}$}
\end{picture}}}
\right]_{\mathrm{eik.}}\right\}_{\mathrm{coll.}} 
\nl&=&
\sum_{\varphi_{i''_k}}
\vcenter{\hbox{\begin{picture}(80,60)(0,-30)
\Line(55,0)(10,0)
\GCirc(55,0){15}{1}
\Text(10,5)[lb]{\scriptsize$\varphi_{i''_k}$}
\end{picture}}}
\de^\coll_{\varphi_{i''_k}\varphi_{i_k}},
\eeqar
where the curly bracket,
consisting of truncated (trunc.) diagrams and
subtracted eikonal contributions (eik.), is evaluated in collinear
approximation (coll.).  
The sum over $V^a$ extends over $\PWp$ and $\PWm$, although in many
cases only one of them contributes.  
The detailed proof of 
\refeq{collfactorization} depends on the spin of the external
particles, which may be scalar bosons ($\varphi_i=\Phi_i$), transverse
gauge bosons ($\varphi_i=V^a_\rT$), or fermions
($\varphi_i=\Psi^\kappa_{j,\si}$).  However, its basic structure can
be sketched in a universal way and consists of two main steps:
\begin{itemize}
\item After insertion of the expressions for the explicit vertices and
  propagators, explicit subtraction of the eikonal contributions, and in
  the limit of collinear gauge-boson emission, the l.h.s.\ of
  \refeq{collfactorization} turns into\footnote{Here and in the
    following the $+\ri\varepsilon$ prescription of the propagators is
    suppressed in the notation.}
\beqar \label{UNIVcollamp}
\lefteqn{
\delta^\coll \M^{\varphi_{i_k}}(p_k)= \sum_{V^a=A,Z,W^\pm} \sum_{\varphi_{i'_k}} \mu^{4-D}\int\ddq
\frac{-\ri  e I^{\bar{V}^a}_{\varphi_{i'_k}\varphi_{i_k}}}{(q^2-M_{V^a}^2)[(p_k-q)^2-M_{\varphi_{i'_k}}^2]} \,K_{\varphi_{i_k}}} \quad
\nl&&\times \lim_{q^\mu\rightarrow xp_k^\mu} 
q^\mu
\left\{
\vcenter{\hbox{\begin{picture}(85,80)(12,20)
\Line(55,60)(15,60)
\Text(35,65)[b]{\scriptsize $\varphi_{i'_k}(p_k-q)$}
\Photon(30,30)(70,60){1}{6}
\Text(55,35)[t]{\scriptsize $V^a_\mu(q)$}
\GCirc(70,60){15}{1}
\end{picture}}}
- \sum_{\varphi_j}
\vcenter{\hbox{\begin{picture}(106,80)(-18,20)
\Line(55,60)(35,60)
\Line(35,60)(0,60)
\Text(30,65)[br]{\scriptsize $ \varphi_{i'_k}(p_k-q)$}
\Text(33,68)[bl]{\scriptsize $ \varphi_{j}(p_k)$}
\Photon(10,30)(35,60){1}{5}
\Vertex(35,60){2}
\Text(35,40)[t]{\scriptsize $V^a_\mu(q)$}
\GCirc(70,60){15}{1}
\end{picture}}}
\right\},
\hspace{2cm}
\eeqar     
where $\ri eI^{V^a}_{\varphi_{i'}\varphi_{i}}$ 
is the coupling
corresponding to the $V^a\bar{\varphi}_{i'}\varphi_{i}$ vertex with
all fields incoming.  The $I^{{V}^a}$ are the the generators of
$\SUtwo\times\Uone$ transformations of the fields $\varphi_{i_k}$ and
are discussed in detail in App.~B of \citere{Denner:2001jv}.  The
charge-conjugate fields are denoted by $\bar\varphi_{i_k}$.  For
scalar bosons and transverse gauge bosons $K_{\varphi_{i_k}}=1$, while
for fermions $K_{\varphi_{i_k}}=2$. The first diagram appearing in
\refeq{UNIVcollamp} results from the first diagram of
\refeq{collfactorization} by omitting the explicit vertex and
propagators. The second diagram in \refeq{UNIVcollamp} originates from
the truncation of the self-energy and mixing-energy
($\varphi_i \varphi_j)$ insertions in the first diagram of
\refeq{collfactorization}.  Equation \refeq{UNIVcollamp} is derived in
\refses{se:fac_scal}--\ref{se:fac_fer}.

\item The contraction of the diagrams between the curly brackets on
  the r.h.s.\ of \refeq{UNIVcollamp} with the gauge-boson momentum
  $q^\mu$ can be simplified using the Ward identities
\beqar\label{UNIVcollwi}
\lefteqn{\lim_{q^\mu\rightarrow xp_k^\mu} q^\mu
\left\{
\vcenter{\hbox{\begin{picture}(85,80)(12,20)
\Line(55,60)(15,60)
\Text(35,65)[b]{\scriptsize $\varphi_{i'_k}(p_k-q)$}
\Photon(30,30)(70,60){1}{6}
\Text(55,35)[t]{\scriptsize $V^a_\mu(q)$}
\GCirc(70,60){15}{1}
\end{picture}}}
- \sum_{\varphi_j}
\vcenter{\hbox{\begin{picture}(100,80)(-10,20)
\Line(55,60)(35,60)
\Line(35,60)(0,60)
\Text(30,65)[br]{\scriptsize $ \varphi_{i'_k}(p_k-q)$}
\Text(33,68)[bl]{\scriptsize $ \varphi_{j}(p_k)$}
\Photon(10,30)(35,60){1}{5}
\Vertex(35,60){2}
\Text(35,40)[t]{\scriptsize $V^a_\mu(q)$}
\GCirc(70,60){15}{1}
\end{picture}}}
\right\}}\quad&&\hspace{10cm}
\nl&&\quad
=\sum_{\varphi_{i_k''}}
\vcenter{\hbox{\begin{picture}(80,80)(0,-40)
\Line(55,0)(10,0)
\GCirc(55,0){15}{1}
\Text(10,5)[lb]{\scriptsize $\varphi_{i_k''}(p_k)$}
\end{picture}}}
 e I^{V^a}_{\varphi_{i_k''}\varphi_{i'_k}},
\eeqar
which are fulfilled in the collinear approximation and valid
up to mass-suppressed terms. These Ward identities
are derived in \refse{CWIsection} using the BRS invariance of the
spontaneously broken $\SUtwo\times\Uone$ Lagrangian.
\end{itemize}
Combining \refeq{UNIVcollwi} with \refeq{UNIVcollamp}, we obtain
\refeq{collfactorization}  with the collinear factor
\beqar\label{loopLA}
\de^\coll_{\varphi_{i''}\varphi_{i}}
&=&
\sum_{V^a=A,Z,W^\pm}\sum_{\varphi_{i'}}
\mu^{4-D}\int\ddq
\frac{-\ri K_{\varphi_{i}} e^2I^{V^a}_{\varphi_{i''}\varphi_{i'}}I^{\bar{V}^a}_{\varphi_{i'}
\varphi_{i}}}{(q^2-M_{V^a}^2)[(p-q)^2-M_{\varphi_{i'}}^2]}
\nl&\LA&
\frac{\alpha}{4\pi}K_{\varphi_{i}}\left[\cew_{\varphi_{i''}\varphi_{i}}\log{\frac{\mu^2}{\MW^2}}+ \de_{\varphi_{i''}\varphi_{i}}Q^2_{\varphi_{i}}\log{\frac{\MW^2}{M_{\varphi_i}^2}} \right],
\eeqar
where the effective electroweak 
Casimir operator is defined by 
\beq\label{Casimirdef}
\sum_{V^a=A,Z,W^\pm}\sum_{\varphi_{i'}}
I^{V^a}_{\varphi_{i''}\varphi_{i'}}
I^{\bar{V}^a}_{\varphi_{i'}\varphi_{i}}
=\cew_{\varphi_{i''}\varphi_{i}}
\eeq 
and explicitly given in App.~B of \citere{Denner:2001jv}.
The integral is evaluated in \refapp{app:masssing}. For virtual massive gauge
bosons $V^a=Z,W^\pm$,
the scale of the logarithm is determined by $\MW$, for photons
by the mass $M_{\varphi_i}$ of the external particles.

In this section, we use the collinear Ward identities 
to derive \refeq{collfactorization} for 
external scalars, transverse gauge bosons, and fermions.
To this end, we introduce the 
following shorthand notation for matrix 
elements \refeq{Bornampli}
\beq \label{shortBornampli}
\M^{\varphi_{i_k}}(p_{k})= v_{\varphi_{i_k}}(p_{k})G^{\underline{\varphi}_{i_k}\underline{\Oper}}(p_{k}),
\eeq
\ie we concentrate on a specific external leg $\varphi_{i_k}$, and
only its momentum $p_{k}$ and wave function $v_{\varphi_{i_k}}(p_k)$
are kept explicit.  
The wave function $v_{\varphi_{i_k}}(p_k)$ equals 1 for scalars and is
given by the Dirac-spinors for fermions and the polarization vectors
for gauge bosons.  It is contracted with the truncated Green function
$G$ (underlined field arguments correspond to truncated external legs,
other conventions concerning Green functions are given in
\refapp{app:GFs}).  The operator
\beq
\Oper(r)=\prod_{l\neq k}\varphi_{i_l}(p_l),\qquad r=\sum_{l\neq k}p_l,
\eeq
represents the remaining external legs.
The external lines corresponding to the operator
$O$ are always assumed to be on-shell and contracted with the
corresponding wave functions. These wave functions are always 
suppressed in the notation. Moreover, often also 
the operator $\Oper$ and the corresponding total 
momentum $r$ are not written.

Note that in intermediate results, owing to gauge-boson emission $\varphi_{i_k} \rightarrow V^a \varphi_{i'_k}$, the matrix elements \refeq{shortBornampli} are modified into expressions where the wave function $v_{\varphi_{i_k}}(p_{k})$ with mass $p_k^2=M_{\varphi_{i_k}}^2$ is contracted with a line $\varphi_{i'_k}$ carrying a different mass  $M_{\varphi_{i'_k}}^2$. In the limit $s\gg
M_{\varphi_{i_k}}^2,M_{\varphi_{i'_k}}^2$, 
the modified matrix elements 
 are identified with matrix elements for $\varphi_{i'_k}$, since
\beq \label{shortBornampli2}
v_{\varphi_{i_k}}(p_{k})G^{\underline{\varphi}_{i'_k}\underline{\Oper}}(p_{k})=
\M^{\varphi_{i'_k}}(p_{k})+\O\left(\frac{M^2}{s}\M^{\varphi_{i'_k}} \right).
\eeq

For the Green functions
corresponding to the diagrams within the curly brackets in
\refeq{UNIVcollamp} we introduce the shorthand
\beqar \label{subtractGF}
G^{[\underline{V}^a\underline{\varphi}_{i}] \underline{\Oper}}_\mu(q,p-q,r)&=&G^{\underline{V}^a\underline{\varphi}_i \underline{\Oper}}_\mu(q,p-q,r) 
- \sum_{\varphi_j}G^{\underline{V}^a \underline{\varphi}_i \varphi_j}_{\mu}(q,p-q,-p)
 G^{\underline{\varphi}_j \underline{\Oper}}(p,r).\nln
\eeqar

\subsection{Factorization for scalars}
\label{se:fac_scal}
We first 
consider the collinear enhancements  generated by the virtual splittings
\beq
\Phi_{i_k}(p_k) \rightarrow V_\mu^a(q) \Phi_{i'_k}(p_k-q),
\eeq
where an incoming on-shell Higgs boson or would-be 
Goldstone boson
$\Phi_{i_k}=H,\chi,\phi^\pm$ emits a virtual collinear gauge boson
$V^a=A,Z,W^\pm$. The corresponding amplitude is given by 
\beqar\label{SCAcollfactorization}
\delta^\coll \M^{\Phi_{i_k}}(p_k)&=&\sum_{V^a} \sum_{\Phi_{i'_k}} \left\{
\left[
\vcenter{\hbox{\begin{picture}(85,60)(0,-30)
\DashLine(5,0)(25,0){4}
\DashLine(25,0)(60.9,14.1){4}
\Photon(25,0)(60.9,-14.1){-2}{3}
\Vertex(25,0){2}
\GCirc(65,0){15}{1}
\Text(35,-14)[lt]{\scriptsize$V^a$}
\Text(35,+12)[lb]{\scriptsize$\Phi_{i'_k}$}
\Text(5,5)[lb]{\scriptsize $\Phi_{i_k}$}
\end{picture}}}
\right]_{\mathrm{trunc.}}
- \sum_{l\neq k}\sum_{\varphi_{i'_l}}\left[
\vcenter{\hbox{\begin{picture}(80,80)(5,-40)
\DashLine(25,30)(39,19.5){4}\DashLine(39,19.5)(53,9){4}
\Line(65,0)(25,-30)
\PhotonArc(65,0)(32.5,143.13,216.87){2}{4.5}
\Vertex(39,19.5){2}
\Vertex(39,-19.5){2}
\GCirc(65,0){15}{1}
\Text(30,0)[r]{\scriptsize$V^a$}
\Text(20,27)[rb]{\scriptsize$\Phi_{i_k}$}
\Text(46,19)[lb]{\scriptsize$\Phi_{i'_k}$}
\Text(20,-27)[rt]{\scriptsize$\varphi_{i_l}$}
\Text(46,-19)[lt]{\scriptsize$\varphi_{i'_l}$}
\end{picture}}}
\right]_{\mathrm{eik.}}\right\}_{\mathrm{coll.}}
\nln
\eeqar
and reads 
\beqar \label{SCAcollamp}
\delta^\coll \M^{\Phi_{i_k}}(p_k)&=& \sum_{V^a=A,Z,W^\pm} \sum_{\Phi_{i'_k}=H,\chi,\phi^\pm} \mu^{4-D}\int\ddq
\frac{\ri e I^{\bar{V}^a}_{\Phi_{i'_k}\Phi_{i_k}}}{(q^2-M_{V^a}^2)[(p_k-q)^2-M_{\Phi_{i'_k}}^2]}
\nl&&\times\lim_{q^\mu\rightarrow xp_k^\mu} \Biggl\{
(2p_k-q)^\mu
G_{\mu}^{[\underline{V}^a\underline{\Phi}_{i'_k}]}(q,p_k-q)
\nl&&\hspace{2cm}
{}+2p_k^\mu \sum_{l\neq k}\sum_{\varphi_{i'_l}}\frac{ 2e p_{l\mu} I^{V^a}_{\varphi_{i'_l}\varphi_{i_l}}  }{[(p_l+q)^2-M_{\varphi_{i'_l}}^2]}
\M^{\Phi_{i'_k}\varphi_{i'_l}}(p_k,p_l)
\Biggr\}.
\eeqar     
According to the definition \refeq{subtractGF}, we have
\beqar\label{SCAampdiag}
\lefteqn{G^{[\underline{V}^a \underline{\Phi}_i]}_{\mu}(q,p-q)=}\quad\nl&&
\vcenter{\hbox{\begin{picture}(90,60)(0,30)
\DashLine(55,60)(10,60){4}
\Text(35,65)[b]{\scriptsize $\Phi_i(p-q)$}
\Photon(30,30)(70,60){2}{6}
\Text(55,35)[t]{\scriptsize $V^a_\mu(q)$}
\GCirc(70,60){15}{1}
\end{picture}}}
-\sum_{\Phi_j}
\vcenter{\hbox{\begin{picture}(95,60)(0,30)
\DashLine(70,60)(35,60){4}
\DashLine(35,60)(5,60){4}
\Text(35,65)[br]{\scriptsize $\Phi_i(p-q)$}
\Text(40,65)[bl]{\scriptsize $\Phi_j(p)$}
\Photon(10,30)(35,60){2}{5}
\GCirc(35,60){2}{0}
\Text(35,40)[t]{\scriptsize $V^a_\mu(q)$}
\GCirc(75,60){15}{1}
\end{picture}}}
-\sum_{V^c}
\vcenter{\hbox{\begin{picture}(95,60)(0,30)
\Photon(70,60)(35,60){2}{3}
\DashLine(35,60)(5,60){4}
\Text(35,65)[br]{\scriptsize $\Phi_i(p-q)$}
\Text(40,65)[bl]{\scriptsize $V^c(p)$}
\Photon(10,30)(35,60){2}{5}
\GCirc(35,60){2}{0}
\Text(35,40)[t]{\scriptsize $V^a_\mu(q)$}
\GCirc(75,60){15}{1}
\end{picture}}}.
\eeqar
Note that the subtracted contributions, when inserted in
\refeq{SCAcollfactorization}, correspond to external scalar
self-energies ($\Phi\Phi$) and scalar--vector mixing-energies ($\Phi
V$).

We first concentrate on the expression between the curly brackets in
\refeq{SCAcollamp}, which has to be evaluated in the collinear limit
$q^\mu \rightarrow x p_k^\mu$. Using
\beq\label{colleik}
\lim_{q^\mu\rightarrow xp_k^\mu}\frac{ 2p_k p_{l}}{[(p_l+q)^2-M_{\varphi_{i'_l}}^2]}=\frac{1}{x}+\O\left(\frac{M^2}{s}\right)
\eeq
and $p_k^\mu \rightarrow q^\mu/x$, one finds%
\footnote{Since the soft-photon contributions are subtracted,
 we do not need a regularization of  $1/x$ for $x\rightarrow 0$.}
\beqar
\lim_{q^\mu\rightarrow xp_k^\mu} \biggl\{\ldots\biggr\} &=&\lim_{q^\mu\rightarrow x p_k^\mu}
\left\{
\left(\frac{2}{x}-1\right)q^\mu
G_{\mu}^{[\underline{V}^a\underline{\Phi}_{i'_k}]}(q,p_k-q)
\right.\nl&&\left.{}
+\frac{2e}{x}\sum_{l\neq k}\sum_{\varphi_{i'_l}} I^{V^a}_{\varphi_{i'_l}\varphi_{i_l}}
\M^{\Phi_{i'_k}\varphi_{i'_l}}(p_k,p_l)
\right\}.\nln
\eeqar
With the  collinear Ward identity \refeq{SCACWIres} for scalar bosons ($\varphi_i=\Phi_{i}$), 
this becomes
\beqar
\lefteqn{\lim_{q^\mu\rightarrow xp_k^\mu} \biggl\{\ldots\biggr\}
=-e\sum_{\Phi_{i''_k}}I^{V^a}_{\Phi_{i''_k}\Phi_{i'_k}}
\M^{\Phi_{i''_k}}(p_k)}\quad\nl
&&{}+
\frac{2e}{x}\left\{\sum_{\Phi_{i''_k}}
I^{V^a}_{\Phi_{i''_k}\Phi_{i'_k}}\M^{\Phi_{i''_k}}(p_k)
+\sum_{l\neq k} \sum_{\varphi_{i'_l}}I^{V^a}_{\varphi_{i'_l}\varphi_{i_l}}
\M^{\Phi_{i'_k}\varphi_{i'_l}}(p_k,p_l)
\right\}.
\eeqar
We now use global $\ewgroup$ invariance, which 
leads to
\beq\label{globinv}
\ri e \sum_{k=1}^n\sum_{\varphi_{i'_k}} I^{V^a}_{\varphi_{i'_k}\varphi_{i_k}} 
\M^{\varphi_{i_1} \ldots \varphi_{i'_k} \ldots \varphi_{i_n}}
=\O\left(\frac{M^2}{s} \M^{\varphi_{i_1} \ldots \varphi_{i_k} \ldots \varphi_{i_n}}\right)
\eeq
for non-mass-suppressed matrix elements.  With this, the part
proportional to $1/x$ is mass-suppressed as expected, since the
soft-photon contributions have been subtracted.  Thus,
\refeq{SCAcollamp} turns into
\beqar
\delta^\coll \M^{\Phi_{i_k}}(p_k) &=&  \sum_{V^a,\Phi_{i'_k},\Phi_{i''_k}}\mu^{4-D}\int\ddq
\frac{-\ri e^2 I^{V^a}_{\Phi_{i''_k}\Phi_{i'_k}}I^{\bar{V}^a}_{\Phi_{i'_k}\Phi_{i_k}}}{(q^2-M_{V^a}^2)[(p_k-q)^2-M_{\Phi_{i'_k}}^2]}
\M^{\Phi_{i''_k}}(p_k),\nl
\eeqar   
and with \refeq{loopLA} and $M_{\Phi_{i'_k}}\sim\MW$ 
we obtain the collinear factor
\beq \label{scalarcollfact}
\de^\coll_{\Phi_{i''}\Phi_{i}}\LA
\frac{\alpha}{4\pi}\de_{\Phi_{i''}\Phi_{i}}\cew_{\Phi}\log{\frac{\mu^2}{\MW^2}}
\eeq
in LA.  For external Higgs bosons, this has to be combined with the
Higgs FRC \cite{Denner:2001jv} as in \refeq{subllogfact2}. The
resulting collinear correction factor reads
\beq\label{Higgscc} 
\de^{\cc}_{HH}=
\frac{\alpha}{4\pi}\left[2\cew_\Phi-\frac{3}{4\sw^2}\frac{\Mt^2}{\MW^2}\right]\log{\frac{s}{\MW^2}},
\eeq 
where $\sw$ represents the sine of the weak mixing angle.
The collinear correction factors for external longitudinal gauge
bosons are obtained from \refeq{scalarcollfact} and from the
corrections to the equivalence theorem \cite{Denner:2001jv}, and read
\beq \label{longeq:coll} 
\delta^\cc_{V_\rL^{b''}V_\rL^{b}}=\delta_{V^{b''}V^{b}}
\frac{\alpha}{4\pi}\left\{\left[2\cew_\Phi-\frac{3}{4\sw^2}\frac{\Mt^2}{\MW^2}\right]\log{\frac{s}{\MW^2}}   
+Q_{V^b}^2\log{\frac{\MW^2}{\la^2}}\right\}.
\eeq
As pointed out in \citere{Denner:2001jv}, Higgs bosons and
longitudinal gauge bosons receive the same collinear SL corrections.
The difference between \refeq{Higgscc} and \refeq{longeq:coll}
consists only in an electromagnetic soft contribution, which is
contained in the FRC for charged gauge bosons and depends on the
infinitesimal photon mass $\la$.  This suggests that the
logarithmic corrections for longitudinal gauge bosons can be
reproduced in the symmetric approach.  In fact, the result
\refeq{longeq:coll} is equivalent to
\beq \label{longeq:coll2} 
\delta^\cc_{V_\rL^{b'}V_\rL^{b}}\LA \left.\left(\frac{1}{2}\delta Z_{\Phi_{b'}\Phi_{b}}
+ \delta^\coll_{\Phi_{b'}\Phi_{b}}\right)\right|_{\mu^2=s},
\eeq
if a FRC for on-shell would-be Goldstone bosons
\beq
\delta Z_{\Phi_{b'}\Phi_{b}}=\left.\left(\frac{\partial}{\partial p^2} \Sigma_{\Phi_{b'}\Phi_{b}}(p^2)\right)\right|_{p^2=M^2_{V^b}},
\eeq
is used where, however, the contributions from $\mathcal{L}_{v}$ have
to be omitted.  Note that \refeq{longeq:coll2} corresponds to the
collinear factor for physical scalar bosons belonging to a Higgs
doublet with vanishing vev.  This result can be interpreted as
follows: as far as logarithmic one-loop corrections are concerned, at
high energies the longitudinal gauge bosons can be described by
would-be Goldstone bosons as physical scalar bosons.  This justifies
the symmetric approach at the one-loop leading-logarithmic level.

\subsection{Factorization for transverse gauge bosons}
\label{se:fac_vec}
Next, we consider the collinear enhancements generated by the virtual
splittings
\beq\label{GAUGEsplitting}
V^{b_k}_\nu(p_k) \rightarrow V_\mu^a(q) V^{b'_k}_{\nu'}(p_k-q),
\eeq
where an incoming on-shell transverse gauge boson $V_\rT^{b_k}=A_\rT,Z_\rT,W_\rT^\pm$ 
emits a virtual collinear gauge boson $V^a=A,Z,W^\pm$.
The corresponding amplitude is given by 
\beqar\label{GAUGEcollfactorization}
\delta^\coll \M^{V_\rT^{b_k}}(p_k)&=&\frac{1}{2}\sum_{V^a,V^{b'_k}} \left\{\left[
\vcenter{\hbox{\begin{picture}(85,60)(0,-30)
\Photon(5,0)(25,0){2}{2}
\Photon(25,0)(60.9,14.1){2}{3}
\Photon(25,0)(60.9,-14.1){-2}{3}
\Vertex(25,0){2}
\GCirc(65,0){15}{1}
\Text(35,-14)[lt]{\scriptsize$V^a$}
\Text(35,+12)[lb]{\scriptsize$V^{b'_k}$}
\Text(5,5)[lb]{\scriptsize $V_\rT^{b_k}$}
\end{picture}}}
\right]_{\mathrm{trunc.}}
\right.\nl&&\left.
{}- \sum_{l\neq k}\sum_{\varphi_{i'_l}}\left[
\vcenter{\hbox{\begin{picture}(80,80)(5,-40)
\Photon(25,30)(53,9){2}{4.5}
\Line(65,0)(25,-30)
\PhotonArc(65,0)(32.5,143.13,216.87){2}{4.5}
\Vertex(39,19.5){2}
\Text(46,19)[lb]{\scriptsize$V^{b'_k}$}
\Text(46,-19)[lt]{\scriptsize$\varphi_{i'_l}$}
\Vertex(39,-19.5){2}
\GCirc(65,0){15}{1}
\Text(30,0)[r]{\scriptsize$V^a$}
\Text(20,27)[rb]{\scriptsize$V_\rT^{b_k}$}
\Text(20,-27)[rt]{\scriptsize$\varphi_{i_l}$}
\end{picture}}}
+
\vcenter{\hbox{\begin{picture}(80,80)(5,-40)
\Photon(25,30)(53,9){2}{4.5}
\Line(65,0)(25,-30)
\PhotonArc(65,0)(32.5,143.13,216.87){2}{4.5}
\Vertex(39,19.5){2}
\Text(46,19)[lb]{\scriptsize$V^{a}$}
\Text(46,-19)[lt]{\scriptsize$\varphi_{i'_l}$}
\Vertex(39,-19.5){2}
\GCirc(65,0){15}{1}
\Text(30,0)[r]{\scriptsize$V^{b'_k}$}
\Text(20,27)[rb]{\scriptsize$V_\rT^{b_k}$}
\Text(20,-27)[rt]{\scriptsize$\varphi_{i_l}$}
\end{picture}}}
\right]_{\mathrm{eik.}}\right\}_{\mathrm{coll.}}.  \eeqar  The
r.h.s.\ 
of \refeq{GAUGEcollfactorization} is manifestly
symmetric with respect to an interchange of 
the gauge bosons $V^a$ and $V^{b'_k}$
resulting from the splitting \refeq{GAUGEsplitting}.  In particular,
the subtracted eikonal contributions are decomposed into terms
originating from soft $V^a$ bosons ($q^\mu\rightarrow 0$) as well as
from soft $V^{b'_k}$ bosons ($q^\mu\rightarrow p_k^\mu$).  The
symmetry factor $1/2$ compensates double counting in the sum over $V^a,
V^{b'_k}=A,Z,W^\pm$.  The resulting amplitude is
\newcommand{\GBvertex}{F}
\beqar \label{GAUGEcollamp}
\lefteqn{\delta^\coll \M^{V_\rT^{b_k}}(p_k)= \frac{1}{2}\sum_{V^a,V^{b'_k}} \mu^{4-D}\int\ddq
\frac{\ri e I^{\bar{V}^a}_{V^{b'_k}V^{b_k}}}{(q^2-M_{V^a}^2)[(p_k-q)^2-M_{V^{b'_k}}^2]}}\quad
\nl&&\times\lim_{q^\mu\rightarrow xp_k^\mu}\varepsilon_{\rT\nu}(p_k) \Biggl\{
\GBvertex^{\mu\nu\nu'}(q,p_k-q)
G_{\mu\nu'}^{[\underline{V}^a\underline{V}^{b'_k}]}(q,p_k-q)
\nl&&
\quad {}+\sum_{l\neq k}\sum_{\varphi_{i'_l}}
\Biggl[
\GBvertex^{\mu\nu\nu'}(0,p_k)
\frac{ 2e p_{l\mu} I^{V^a}_{\varphi_{i'_l}\varphi_{i_l}}  }{[(p_l+q)^2-M_{\varphi_{i'_l}}^2]}
G_{\nu'}^{\underline{V}^{b'_k}\underline{\varphi}_{i'_l}}(p_k,p_l)
\nl&&
\quad {}+\GBvertex^{\mu\nu\nu'}(p_k,0)
\frac{ 2e p_{l\nu'} I^{V^{b'_k}}_{\varphi_{i'_l}\varphi_{i_l}}  }{[(p_l+p_k-q)^2-M_{\varphi_{i'_l}}^2]}
G_{\mu}^{\underline{V}^{a}\underline{\varphi}_{i'_l}(p_l)}(p_k,p_l)
\Biggr]v_{\varphi_{i_l}(p_l)}\Biggr\},
\eeqar     
where  
\beq\label{3Vcoup}
\GBvertex^{\mu\nu\nu'}(q,p_k-q)=
\left[ 
g^{\nu\nu'}(2p_k-q)^\mu+g^{\nu'\mu}(2q-p_k)^\nu-g^{\mu\nu}(p_k+q)^{\nu'}\right]
\eeq
is the vertex function associated to the splitting
\refeq{GAUGEsplitting}.  According to the definition
\refeq{subtractGF}, 
\beqar\label{GAUGEampdiag}
\lefteqn{G^{[\underline{V}^a \underline{V}^b]}_{\mu\nu}(q,p-q)=}\quad\nl&&
\vcenter{\hbox{\begin{picture}(90,60)(0,30)
\Photon(55,60)(10,60){2}{6}
\Text(30,65)[b]{\scriptsize $V^b_\nu(p-q)$}
\Photon(30,30)(70,60){2}{6}
\Text(55,35)[t]{\scriptsize $V^a_\mu(q)$}
\GCirc(70,60){15}{1}
\end{picture}}}
-\sum_{V^c}
\vcenter{\hbox{\begin{picture}(95,60)(0,30)
\Photon(70,60)(35,60){2}{4}
\Photon(35,60)(5,60){2}{4}
\Text(35,65)[br]{\scriptsize $V^b_\nu(p-q)$}
\Text(40,65)[bl]{\scriptsize $V^c(p)$}
\Photon(10,30)(35,60){2}{5}
\GCirc(35,60){2}{0}
\Text(35,40)[t]{\scriptsize $V^a_\mu(q)$}
\GCirc(75,60){15}{1}
\end{picture}}}
-\sum_{\Phi_j}
\vcenter{\hbox{\begin{picture}(95,60)(0,30)
\DashLine(70,60)(35,60){3}
\Photon(35,60)(5,60){2}{4}
\Text(35,65)[br]{\scriptsize $V^b_\nu(p-q)$}
\Text(40,65)[bl]{\scriptsize $\Phi_j(p)$}
\Photon(10,30)(35,60){2}{5}
\GCirc(35,60){2}{0}
\Text(35,40)[t]{\scriptsize $V^a_\mu(q)$}
\GCirc(75,60){15}{1}
\end{picture}}}.
\eeqar

We first concentrate on the contraction of the vertex \refeq{3Vcoup}
with the transverse polarization vector $\varepsilon^\nu_\rT(p_k)$.
Owing to $p_{k}^\nu\varepsilon_{\rT\nu}(p_k)=0$, 
the second term on the
r.h.s.\ of \refeq{3Vcoup} vanishes in the collinear limit $q^\mu
\rightarrow x p_k^\mu$, and
\beq\label{GAUGEcollvert}
\lim_{q^\mu\rightarrow xp_k^\mu}\varepsilon_{\rT\nu}(p_k)\GBvertex^{\mu\nu\nu'}(q,p_k-q)
=
\left(\frac{2}{x}-1\right)\varepsilon^{\nu'}_\rT(p_k)q^\mu
-\left(\frac{2}{1-x}-1\right)\varepsilon^{\mu}_\rT(p_k)(p_k-q)^{\nu'}.
\eeq
In the fractions $2/x$ and $2/(1-x)$ we have isolated the terms
leading to IR enhancements at $x\rightarrow 0$ and $x\rightarrow 1$,
respectively.  These must be cancelled by the subtracted eikonal
contributions, \ie the terms in the last two lines 
in \refeq{GAUGEcollamp}.  In these contributions some terms are
mass-suppressed or vanishing owing to Ward identities for the massive
or massless on-shell gauge bosons $V^{b'_k}$ and $V^a$, respectively,
\beqar
\varepsilon^{\mu}_\rT(p_k) p_{l\mu} p_k^{\nu'}
G_{\nu'}^{\underline{V}^{b'_k}\underline{\varphi}_{i'_l}}(p_k,p_l)&\sim&
M_{V^{b'_k}} \M^{V_\rL^{b'_k}{\varphi}_{i'_l}}(p_k,p_l), \nl
\varepsilon^{\nu'}_\rT(p_k) p_{l\nu'} p_k^\mu
G_{\mu}^{\underline{V}^a\underline{\varphi}_{i'_l}}(p_k,p_l)&\sim&
M_{V^a} \M^{V_\rL^a{\varphi}_{i'_l}}(p_k,p_l).
\eeqar
Thus, the relevant terms are obtained by the substitutions
\beqar\label{GAUGEeikvert}
\varepsilon_{\rT\nu}(p_k) \GBvertex^{\mu\nu\nu'}(0,p_k)&\to& 
2\varepsilon^{\nu'}_\rT(p_k)p_k^{\mu},
\nl
\varepsilon_{\rT\nu}(p_k) \GBvertex^{\mu\nu\nu'}(p_k,0)&\to& 
-2\varepsilon^{\mu}_\rT(p_k)p_k^{\nu'}
\eeqar
in \refeq{GAUGEcollamp}.
With \refeq{GAUGEcollvert} and \refeq{GAUGEeikvert}, the expression
between the curly brackets on the r.h.s.\ of \refeq{GAUGEcollamp}
gives
\beqar \label{GAUGEcollamp2}
\lefteqn{\lim_{q^\mu\rightarrow xp_k^\mu} \biggl\{\ldots\biggr\} =}\quad
\nl&=&{}
\lim_{q^\mu\rightarrow xp_k^\mu}
\biggl\{
\left[\left(\frac{2}{x}-1\right)\varepsilon^{\nu'}_\rT(p_k)q^\mu
-\left(\frac{2}{1-x}-1\right)\varepsilon^{\mu}_\rT(p_k)(p-q)^{\nu'}
\right]
G_{\mu\nu'}^{[\underline{V}^a\underline{V}^{b'_k}]}(q,p_k-q)
\nl&&{}
+\sum_{l\neq k}\sum_{\varphi_{i'_l}}
\biggl[
\frac{2e}{x} I^{V^a}_{\varphi_{i'_l}\varphi_{i_l}}
\M^{V_\rT^{b'_k}\varphi_{i'_l}}(p_k,p_l)
-\frac{2e}{1-x}I^{V^{b'_k}}_{\varphi_{i'_l}\varphi_{i_l}} 
\M^{V_\rT^{a}\varphi_{i'_l}}(p_k,p_l)
\biggr]\biggr\}.
\eeqar
Using  the collinear Ward identity \refeq{SCACWIres}
for gauge bosons ($\varphi_i=V^b_\nu$)
and the equivalent identity 
\beq\label{GAUGECWIres2} 
\lim_{q^\mu \rightarrow xp^\mu}\varepsilon_\rT^\mu(p) (p-q)^{\nu}
 G^{[\underline{V}^a \underline{V}^{b}] \underline{\Oper}}_{\mu\nu}(q,p-q,r)
= e \sum_{V^{b'}} \M^{V_\rT^{b'} \Oper}(p,r) I^{V^{b}}_{V^{b'}V^a},
\eeq
\refeq{GAUGEcollamp2} simplifies into
\beqar \label{GAUGEfacteqb}
\lefteqn{\lim_{q^\mu\rightarrow xp_k^\mu} \biggl\{\ldots\biggr\} = 
-e\sum_{V^{b''_k}} \left[I^{V^a}_{V^{b''_k}V^{b'_k}}-I^{V^{b'_k}}_{V^{b''_k}V^{a}}\right]\M^{V_\rT^{b''_k}}(p_k)}\quad
\nl&&{}
+
\frac{2e}{x}\biggl(\sum_{V^{b''_k}} I^{V^a}_{V^{b''_k}V^{b'_k}}\M^{V_\rT^{b''_k}}(p_k)
+ \sum_{l\neq k}\sum_{\varphi_{i'_l}}I^{V^a}_{\varphi_{i'_l}\varphi_{i_l}}
\M^{V_\rT^{b'_k}\varphi_{i'_l}}(p_k,p_l)\biggr)
\nl&&
{}-\frac{2e}{1-x}\biggl(\sum_{V^{b''_k}} I^{V^{b'_k}}_{V^{b''_k}V^{a}} \M^{V_\rT^{b''_k}}(p_k)
+\sum_{l\neq k}\sum_{\varphi_{i'_l}}
I^{{V}^{b'_k}}_{\varphi_{i'_l}\varphi_{i_l}} 
\M^{V_\rT^{a}\varphi_{i'_l}}(p_k,p_l)
\biggr).
\eeqar  
Again, the soft terms proportional to $1/x$ and $1/(1-x)$ 
are mass-suppressed owing to global $\ewgroup$ invariance
\refeq{globinv}, so that only the first term in \refeq{GAUGEfacteqb}
remains. Inserting this into \refeq{GAUGEcollamp} with
$I^{V^{b}}_{V^{c}V^{a}}=-I^{V^a}_{V^{c}V^{b}}$, we find
\beqar
\delta^\coll \M^{V_\rT^{b_k}}(p_k) &=&
\sum_{V^a,V^{b'_k},V^{b''_k}}\mu^{4-D}\int\ddq
\frac{-\ri e^2 I^{V^a}_{V^{b''_k}V^{b'_k}}I^{\bar{V}^a}_{V^{b'_k}V^{b_k}}}{(q^2-M_{V^a}^2)[(p_k-q)^2-M_{V^{b'_k}}^2]}
\M^{V_\rT^{b''_k}\Oper}(p_k).\nln
\eeqar   
With \refeq{loopLA} we obtain the collinear factor
\beq    
\de^\coll_{V_\rT^{b''_k}V_\rT^{b_k}}\LA
\frac{\alpha}{4\pi}\cew_{V^{b''_k}V^{b_k}}\log{\frac{\mu^2}{\MW^2}}
\eeq
in LA.  The complete SL collinear (and soft) correction factors for
transverse gauge bosons are obtained by including the corresponding
FRC's given in \citere{Denner:2001jv} and read
\beq \label{deccWT}
\delta^\cc_{V_\rT^aV_\rT^b}=\frac{\alpha}{4\pi}\left\{\frac{1}{2}\left[\bew_{V^aV^b}+E_{V^aV^b}\bew_{AZ}\right]\log{\frac{s}{\MW^2}} 
+\de_{V^aV^b} Q_{V^a}^2\log{\frac{\MW^2}{\la^2}}\right\}
-\frac{1}{2}\de_{V^a A}\de_{V^bA} \Delta \alpha (\MW^2).
\eeq
The $s$-dependent part is determined by the one-loop coefficients
$\bew_{V^aV^b}$ of the electroweak $\beta$-function (see
\citere{Denner:2001jv}). The remaining terms represent a soft
contribution proportional to the charge of the gauge boson and a pure
electromagnetic contribution originating from light fermion loops that
can be related to the running of the electromagnetic coupling from zero
to the scale $\MW$ (defined explicitly in \citere{Denner:2001jv}).

\subsection{Factorization for fermions}
\label{se:fac_fer}
We finally 
consider the collinear enhancements  generated by the virtual splittings 
\beq
f^\kappa_{j,\si}(p_k) \rightarrow V_\mu^a(q) f^\kappa_{j',\si'}(p_k-q),
\eeq
where a virtual
collinear gauge boson $V^a=A,Z,W^\pm$ is emitted by an
incoming on-shell fermion $f^\kappa_{j,\si}$, \ie a quark or lepton
$f=Q,L$, with chirality $\kappa=\rL,\rR$, isospin index $\si=\pm$, and
generation index $j=1,2,3$.  The collinear singularity is contained in
\beqar\label{FERcollfactorization}
\delta^\coll \M^{f^\kappa_{j,\si}}(p_k)&=&
\sum_{V^a}\sum_{j'\si'}
\left\{
\left[\vcenter{\hbox{\begin{picture}(85,60)(0,-30)
\ArrowLine(5,0)(25,0)
\ArrowLine(25,0)(60.9,14.1)
\Photon(25,0)(60.9,-14.1){-2}{3}
\Vertex(25,0){2}
\GCirc(65,0){15}{1}
\Text(35,-14)[lt]{\scriptsize$V^a$}
\Text(5,5)[lb]{\scriptsize $\Psi_{j,\si}$}
\Text(35,+12)[lb]{\scriptsize$\Psi_{j',\si'}$}
\end{picture}}}
\right]_{\mathrm{trunc.}}
- \sum_{l\neq k}\sum_{\varphi_{i'_l}}\left[
\vcenter{\hbox{\begin{picture}(80,80)(5,-40)
\ArrowLine(25,30)(39,19.5)\ArrowLine(39,19.5)(53,9)
\Line(65,0)(25,-30)
\PhotonArc(65,0)(32.5,143.13,216.87){2}{4.5}
\Vertex(39,19.5){2}
\Vertex(39,-19.5){2}
\GCirc(65,0){15}{1}
\Text(30,0)[r]{\scriptsize$V^a$}
\Text(20,27)[rb]{\scriptsize$\Psi_{j,\si}$}
\Text(20,-27)[rt]{\scriptsize$\varphi_{i_l}$}
\Text(46,19)[lb]{\scriptsize$\Psi_{j',\si'}$}
\Text(46,-19)[lt]{\scriptsize$\varphi_{i'_l}$}
\end{picture}}}
\right]_{\mathrm{eik.}}\right\}_{\mathrm{coll.}}.\nln
\eeqar
The corresponding amplitude  reads 
\beqar \label{FERcollamp}
\lefteqn{\delta^\coll \M^{f^\kappa_{j,\si}}(p_k)= \sum_{V^a=A,Z,W^\pm} \sum_{j',\si'} \mu^{4-D}\int\ddq
\frac{\ri e I^{\bar{V}^a}_{\si'\si}U^{\bar{V}^a}_{j'j}}{(q^2-M_{V^a}^2)[(p_k-q)^2-m_{f_{j',\si'}}^2]}}\quad
\nl&&\times\lim_{q^\mu\rightarrow xp_k^\mu} \Biggl\{\Biggl[
G_{\mu}^{[\underline{V}^a\underline{\Psi}_{j',\si'}^\kappa]}(q,p_k-q)(\ps_k-\qs)
\nl&&
{}+\sum_{l\neq k}\sum_{\varphi_{i'_l}}\frac{ 2e p_{l\mu} I^{V^a}_{\varphi_{i'_l}\varphi_{i_l}}  }{[(p_l+q)^2-M_{\varphi_{i'_l}}^2]}
G^{\underline{\Psi}_{j',\si'}^\kappa\underline{\varphi}_{i'_l}}(p_k,p_l)
v_{\varphi_{i_l}}(p_l)\,
\ps_k\Biggr]\gamma^\mu u(p_k)
\Biggr\},\hspace{2cm}
\eeqar   
where the fermion-mass terms in the numerator have been neglected, and
the unitary mixing matrix $U^{V^a}$ is defined in
\refeq{mixingmatrix}.  According to the definition \refeq{subtractGF}
the Green function
$G^{[\underline{V}^a\underline{\Psi}^\kappa_{j,\si}]}_{\mu}$ 
is diagrammatically given by
\beqar\label{FERampdiag}
G^{[\underline{V}^a \underline{\Psi}_{j,\si}^\kappa]}_{\mu}(q,p-q)=
\vcenter{\hbox{\begin{picture}(90,60)(0,30)
\ArrowLine(10,60)(55,60)
\Text(35,65)[b]{\scriptsize $\Psi^\kappa_{j,\si}(p-q)$}
\Photon(30,30)(70,60){2}{6}
\Text(55,35)[t]{\scriptsize $V^a_\mu(q)$}
\GCirc(70,60){15}{1}
\end{picture}}}
-\sum_{\Psibar}\ 
\vcenter{\hbox{\begin{picture}(95,60)(0,30)
\ArrowLine(35,60)(70,60)
\ArrowLine(5,60)(35,60)
\Text(35,65)[br]{\scriptsize $\Psi^\kappa_{j,\si}(p-q)$}
\Text(40,65)[bl]{\scriptsize $\Psi(p)$}
\Photon(10,30)(35,60){2}{5}
\GCirc(35,60){2}{0}
\Text(35,40)[t]{\scriptsize $V^a_\mu(q)$}
\GCirc(75,60){15}{1}
\end{picture}}}.
\eeqar
In the collinear limit, the expression between the curly brackets in
\refeq{FERcollamp} can be simplified using \refeq{colleik},
\beq\label{masslessdirac}
\lim_{q^\mu \rightarrow xp_k^\mu}(\ps_k-\qs) \gamma^\mu  u(p_k)= 
\left(\frac{2}{x}-2\right)q^\mu u(p_k) + \O(m_{j,\si})u(p_k),
\eeq
and the collinear Ward identity \refeq{FERCWIres}.  One obtains
\beqar
\lefteqn{
\lim_{q^\mu\rightarrow xp_k^\mu} \biggl\{\ldots\biggr\} = }\quad\nl
&=&\lim_{q^\mu\rightarrow x p_k^\mu}
\left(\frac{2}{x}-2\right)q^\mu G_{\mu}^{[\underline{V}^a\underline{\Psi}_{j',\si'}^\kappa]}(q,p_k-q)u(p_k)
\nl&&+{}
\frac{2e}{x}\sum_{l\neq k}\sum_{\varphi_{i'_l}} I^{V^a}_{\varphi_{i'_l}\varphi_{i_l}}
\M^{f_{j',\si'}^\kappa\varphi_{i'_l}}(p_k,p_l)
\nl&=&
-2 e\sum_{j'',\si''}I^{V^a}_{\si''\si'}U^{V^a}_{j''j'} \M^{f_{j'',\si''}^\kappa}(p_k)
\nl&&{}+
\frac{2e}{x}\left\{\sum_{j'',\si''}I^{V^a}_{\si''\si'}U^{V^a}_{j''j'} \M^{f_{j'',\si''}^\kappa}(p_k)
+\sum_{l\neq k} \sum_{\varphi_{i'_l}}I^{V^a}_{\varphi_{i'_l}\varphi_{i_l}}
\M^{f_{j',\si'}^\kappa\varphi_{i'_l}}(p_k,p_l)
\right\}.\quad
\eeqar
Again, the soft-photon contributions proportional to $1/x$ 
are mass-suppressed owing to global gauge invariance \refeq{globinv}.
Thus, only the part 
originating from the $\qs$ term in
\refeq{FERcollamp} contributes, and we find
\beqar 
\delta^\coll\M^{f^\kappa_{j,\si}}(p_k) &=&
\sum_{V^a,j',j'',\si',\si''}\mu^{4-D}\int\ddq \frac{-2\ri e^2
  I^{V^a}_{\si''\si'}U^{V^a}_{j''j'}
  I^{\bar{V}^a}_{\si'\si}U^{\bar{V}^a}_{j'j}}
{(q^2-M_{V^a}^2)[(p_k-q)^2-m_{f_{j',\si'}}^2]}
\M^{f_{j'',\si''}^\kappa}(p_k).\nln 
\eeqar 
Using \refeq{loopLA}, and the unitarity of the mixing matrix,
$\sum_{j'}U^{V^a}_{j''j'}U^{\bar{V}^a}_{j'j}=\de_{j''j}$,
the mixing matrix drops out, and we obtain the collinear factor in LA,
\beq
\de^\coll_{f_{j'',\si''}f_{j,\si}}\LA \de_{j''j}\de_{\si''\si}
\frac{\alpha}{2\pi} \left\{ \cew_{f^\kappa_\si}\log{\frac{\mu^2}{\MW^2}}
  +Q^2_{f_{j,\si}}\log{\frac{\MW^2}{m^2_{f_{j,\si}}}} \right\}.  
\eeq
Adding the FRC for fermions \cite{Denner:2001jv}, we obtain the SL collinear
(and soft) corrections 
\beqar \label{deccfer}
\de^{\cc}_{f^\kappa_{j'',\si''}
  f^\kappa_{j,\si}}&=&\de_{jj''}\de_{\si\si''}\frac{\alpha}{4\pi}\left\{\left[\frac{3}{2}
    \cew_{f^\kappa_\si} -\frac{1}{8\sw^2}\left((1+\delta_{\kappa
        \rR})\frac{m_{f_{j,\si}}^2}{\MW^2}+\delta_{\kappa
        \rL}\frac{m_{f_{{j,-\si}}}^2}{\MW^2}\right)\right]\log{\frac{s}{\MW^2}}
\right.\nl&&\hspace{2cm}\left.
  {}+Q_{f_{j,\si}}^2\left[\frac{1}{2}\log{\frac{\MW^2}{m^2_{f_{j,\si}}}}
    +\log{\frac{\MW^2}{\la^2}} \right] \right\}.
\eeqar 
The Yukawa
contributions are large only for external heavy quarks
$f^\kappa_{j,\si}=\Pt^\rR$, $\Pt^\rL$, and $\Pb^\rL$.  In contrast to
the $\Mt^2$ corrections to the $\rho$ parameter, which are only
related to the (virtual) left-handed $(\Pt,\Pb)$ doublet, logarithmic
Yukawa contributions appear also for (external) right-handed top
quarks.

\section{Collinear Ward identities}\label{CWIsection}

As we have already stressed in \refse{sec:Fact}, the proof of the
factorization identities \refeq{collfactorization} is based on the
collinear Ward identities \refeq{UNIVcollwi}.  In the compact notation
introduced in \refeq{shortBornampli} and
\refeq{subtractGF}, these Ward identities read
\beq \label{CWIs}
\lim_{q^\mu \rightarrow xp^\mu}
q^\mu
v_{\varphi}(p)G^{[\underline{V}^a\underline{\varphi}_i]
  \underline{\Oper}}_{\mu}(q,p-q,r)= e
\sum_{\varphi_{i'}}\M^{\varphi_{i'}
\Oper}(p,r) I^{V^a}_{\varphi_i'\varphi_i}, 
\eeq
A detailed derivation of these identities
is presented in \refse{se:SCACWI}, for external scalars
($\varphi_i=\Phi_i$) and gauge bosons ($\varphi_i=V^a$) and in
\refse{se:FERCWI} for fermions ($\varphi_i=\Psi^\kappa_{j,\si}$). Here
we discuss the most important features and restrictions concerning the
Ward identities \refeq{CWIs}:
\begin{itemize}
\item They are restricted to LO matrix elements. We stress that all
  equations used in this section are only valid in LO.
\item They are realized in the high-energy limit \refeq{Sudaklim},
and in the limit of collinear gauge boson momenta
  $q$ and quasi on-shell external momenta $p$, \ie in the limit where
  $0< p^2,(p-q)^2\ll s$. All these limits have to be taken
  simultaneously.
  The wave function $v_{\varphi}(p)$ corresponds to a particle with
  mass $\sqrt{p^2}$.
\item They are valid only up to mass-suppressed terms, 
 to be precise terms of
  the order $M/\sqrt{s}$ (for fermions) or $M^2/s$ (for bosons)
  with respect to the leading terms appearing in \refeq{CWIs}, where 
$M^2\sim\max(p^2,M_{\varphi_i}^2,M^2_{V^a})$.
  Furthermore, they apply only to matrix elements that are not
  mass-suppressed. In other words, they apply to those matrix elements
  that arise from $\mathcal{L_{\mathrm{symm}}}$ in LO.
\item Their derivation is based on the BRS invariance of a spontaneously
  broken gauge theory (see \refapp{BRStra}). In particular, we used
  only the generic form of the BRS transformations of the fields, the
  form of the gauge-fixing term in an arbitrary `t~Hooft gauge,
\refeq{Gfix}, and the corresponding form of the tree-level
propagators.  Therefore, the result is valid for a general
spontaneously broken
gauge theory, in an  arbitrary `t~Hooft gauge.

\end{itemize}
It is important to observe that the identities \refeq{CWIs} do not
reflect the presence of the non-vanishing vev of the Higgs doublet.
In fact, they are identical to the identities obtained within a
symmetric gauge theory with massless gauge bosons. However,
spontaneous symmetry breaking plays a non-trivial role in ensuring the
validity of \refeq{CWIs}. 
It guarantees the cancellation
of mixing terms between gauge bosons and would-be Goldstone bosons.
In particular, we stress the following:
{\em extra contributions originating from $\L_{v}$ cannot be excluded
  a priori in \refeq{CWIs}}.  In fact, the corresponding
mass-suppressed couplings can in principle give extra leading
contributions if they are enhanced by propagators with small
invariants. We show that no such extra terms are left 
in the final result.
Such terms appear, however, in the derivation of the 
Ward identity for external
would-be 
Goldstone bosons ($\varphi_i=\Phi_i$) as ``extra contributions''
involving gauge bosons
($\varphi_{i'}= V^a$), and in the derivation of the 
Ward identity for external gauge
bosons ($\varphi_i=V^a$) as ``extra contributions'' involving
would-be Goldstone bosons ($\varphi_{i'}=\Phi_i$) [see \refeq{SCAMterm}].
Their cancellation is ensured by Ward identities \refeq{brokenWIs}
relating the electroweak vertex functions that involve explicit
factors with mass dimension.
{\em In other words, the validity of \refeq{CWIs} within a spontaneously
broken gauge theory is a non-trivial consequence of the symmetry of
the full theory.}

In the following, the collinear Ward identities are derived for matrix
elements involving the physical fields of the electroweak
theory. 

\subsection{Scalar bosons and transverse gauge bosons}
\label{se:SCACWI}
The Ward identities for external scalar bosons
$\Phi_i=H,\chi,\phi^\pm$ and transverse gauge bosons $V^b=A,Z,W^\pm$
are of the same form.  Here we derive a generic Ward identity for
external bosonic fields $\varphi_i$ valid for $\varphi_i=\Phi_i$ as
well as $\varphi_i=V^b_\mu$.  In both cases mixing between would-be
Goldstone bosons and gauge bosons has to be taken into
account.\footnote{For external Higgs bosons or photons all mixing
  terms vanish.}  Therefore, we use the symbol $\tilde{\varphi}$ to
denote the mixing partner of $\varphi$, \ie we have
$(\varphi,\tilde{\varphi})=(\Phi,V)$ or
$(\varphi,\tilde{\varphi})=(V,\Phi)$.  The resulting Ward identities
read
\beqar\label{SCACWIdiag}
\lefteqn{
\lim_{q^\mu\rightarrow xp^\mu}
q^\mu \times
\left\{
\vcenter{\hbox{\begin{picture}(85,80)(5,20)
\DashLine(55,60)(10,60){4}
\Text(30,65)[b]{\scriptsize $\varphi_i(p-q)$}
\Photon(30,30)(70,60){1}{6}
\Text(55,35)[t]{\scriptsize $V^a_\mu(q)$}
\GCirc(70,60){15}{1}
\end{picture}}}
-\sum_{\varphi_{i'}}
\vcenter{\hbox{\begin{picture}(90,80)(5,20)
\DashLine(60,60)(35,60){4}
\DashLine(35,60)(10,60){4}
\Text(20,65)[b]{\scriptsize $ \varphi_i(p-q)$}
\Text(40,65)[bl]{\scriptsize $ \varphi_{i'}(p)$}
\Photon(10,30)(35,60){1}{5}
\Vertex(35,60){2}
\Text(35,40)[t]{\scriptsize $V^a_\mu(q)$}
\GCirc(75,60){15}{1}
\end{picture}}}
-\sum_{\tilde{\varphi}_j}
\vcenter{\hbox{\begin{picture}(90,80)(5,20)
\Photon(60,60)(35,60){2}{3.5}
\DashLine(35,60)(10,60){4}
\Text(20,65)[b]{\scriptsize $ \varphi_i(p-q)$}
\Text(40,65)[bl]{\scriptsize $\tilde{\varphi}_j(p)$}
\Photon(10,30)(35,60){1}{5}
\Vertex(35,60){2}
\Text(35,40)[t]{\scriptsize $V^a_\mu(q)$}
\GCirc(75,60){15}{1}
\end{picture}}}
\right\}}\quad&&\hspace{13cm}
\nl&&= e \sum_{\varphi_{i'}} I^{V^a}_{\varphi_{i'}\varphi_i}
\vcenter{\hbox{\begin{picture}(80,80)(10,20)
\DashLine(55,60)(25,60){4}
\Text(40,65)[b]{\scriptsize $\varphi_{i'}(p)$}
\GCirc(70,60){15}{1}
\end{picture}}} + \O\left(M^2E^{d-2}\right),
\eeqar
where $M^2\sim\max(p^2,M_{\varphi_i}^2,M^2_{V^a})$, and $d$ is the
mass dimension of the matrix element $\M^{\varphi_i}$. 
The diagrammatic representation
corresponds to external scalars ($\varphi=\Phi$).  For the proof of
\refeq{SCACWIdiag}, we start from the
BRS invariance (\cf\refapp{BRStra}) 
of the Green function $\langle
\bar{u}^{a}(x)\varphi^+_{i}(y)\Oper(z)\rangle$:
\beqar\label{SCABRSinv}
\langle[\brs \bar{u}^{a}(x)]\varphi^+_{i}(y)\Oper(z)\rangle
-\langle\bar{u}^{a}(x)[\brs \varphi^+_{i}(y)]\Oper(z)\rangle
=\langle\bar{u}^{a}(x)\varphi^+_{i}(y)[\brs \Oper(z)]\rangle.
\eeqar
With  the BRS variations \refeq{antighostbrstra} and \refeq{ccphysbrstra} 
this yields
\newcommand{\linbrs}{X}
\beqar
\lefteqn{\frac{1}{\xi_a}\partial^\mu_{x}\langle\bar{V}^a_\mu(x)\varphi^+_{i}(y)\Oper(z)\rangle
-\ri e v\sum_{\Phi_j=H,\chi,\phi^\pm} I^{V^a}_{H\Phi_j}\langle\Phi_j(x) \varphi^+_{i}(y)\Oper(z)\rangle}\quad
\nl&&{}
+\sum_{V^b=A,Z,W^\pm}\biggl[\linbrs^{V^b}_{\varphi^+_i}
\langle\bar{u}^{a}(x)u^{b}(y)\Oper(z)\rangle
-\ri e\sum_{\varphi_{i'}}
\langle\bar{u}^{a}(x)u^{b}(y)\varphi^+_{i'}(y)\Oper(z)\rangle
I^{V^b}_{\varphi_{i'}\varphi_i}\biggr]
\nl&=&
-\langle\bar{u}^{a}(x)\varphi^+_{i}(y)[\brs \Oper(z)]\rangle
.
\eeqar
Fourier transformation of the variables $(x,y,z)$ to the incoming
momenta $(q,p-q,r)$ ($\partial_{x}^\mu \rightarrow \ri q^\mu$) gives
\beqar \label{PSWISCABOS}
\lefteqn{\frac{\ri}{\xi_a} q^\mu\langle\bar{V}^a_\mu(q)\varphi^+_{i}(p-q)\Oper(r)\rangle
{}-\ri e v\sum_{\Phi_j} I^{V^a}_{H\Phi_j}\langle\Phi_j(q) \varphi^+_{i}(p-q)\Oper(r)\rangle}\quad
\nl&&
{}+\sum_{V^b}\linbrs^{V^b}_{\varphi^+_i}
\langle\bar{u}^{a}(q)u^{b}(p-q)\Oper(r)\rangle
\nl&&{}
-\ri e\sum_{V^b,\varphi_{i'}}
\int \ddl
\langle\bar{u}^{a}(q)u^{b}(l)\varphi^+_{i'}(p-q-l)\Oper(r)\rangle
I^{V^b}_{\varphi_{i'}\varphi_i}
\nl&=&
-\langle\bar{u}^{a}(q)\varphi^+_{i}(p-q)[\brs \Oper(r)]\rangle.
\eeqar 
From now on, the r.h.s.\ is omitted, 
since the BRS variation
of on-shell physical fields does not contribute to physical 
matrix elements.
This can be verified by truncation of the physical external legs
$\Oper(r)$ and contraction with the 
wave functions.  A further simplification concerns the last term on
the l.h.s.\ of \refeq{PSWISCABOS}. This originates from the BRS
variation $\brs \varphi^+_{i}(y)$ of the external scalar or vector 
field and contains an external ``BRS vertex'' connecting the fields
$u^{b}(y)\varphi^+_{i'}(y)$, which we represent by a small box in
\refeq{O1O2picture}.  When we restrict the relation to LO connected
Green functions, this term simplifies into those tree diagrams where
the external ghost line is not connected to the scalar leg of the BRS
vertex by internal vertices,
\beqar\label{O1O2picture}
\vcenter{\hbox{\begin{picture}(90,65)(0,20)
\DashLine(55,62)(10,62){4}
\DashArrowLine(55,58)(10,58){1}
\EBox(8,58)(12,62)
\Text(5,60)[r]{\scriptsize $\brs \varphi_i^+(p-q)$}
\DashArrowLine(30,30)(70,60){1}
\Vertex(30,30){2}
\Text(55,35)[t]{\scriptsize $\bar{u}^{a}(q)$}
\GCirc(70,60){15}{1}
\end{picture}}}
=
\vcenter{\hbox{\begin{picture}(90,65)(0,20)
\DashLine(55,62)(25,62){4}
\EBox(23,58)(27,62)
\Text(40,65)[b]{\scriptsize $\varphi^+_{i'}(p)$}
\DashArrowLine(0,30)(23,58){1}
\Vertex(0,30){2}
\Text(25,35)[t]{\scriptsize $\bar{u}^a(q)$}
\GCirc(70,60){15}{1}
\Text(70,60)[c]{\small$O$}
\end{picture}}}
+\sum_{\Oper_1\neq \Oper}
\vcenter{\hbox{\begin{picture}(85,65)(10,20)
\DashLine(70,75)(25,62){4}
\EBox(23,58)(27,62)
\Text(35,75)[b]{\scriptsize $\varphi^+_{i'}(p+r_2)$}
\Vertex(25,30){2}
\DashArrowLine(25,30)(70,45){1}
\DashArrowLine(70,45)(27,58){1}
\Text(45,30)[t]{\scriptsize $\bar{u}^{a}(q)$}
\GCirc(70,75){10}{1}
\Text(70,75)[c]{\small $\Oper_1$}
\GCirc(70,45){10}{1}
\Text(70,45)[c]{\small $\Oper_2$}
\end{picture}}}.
\eeqar
We will see in the following that the relevant contributions result
only from the first diagram on the r.h.s.\ of \refeq{O1O2picture},
where the ghosts are joined by a propagator and all on-shell legs
$\Oper(r)$ are connected to the leg $\varphi^+_{i'}$, which receives
momentum $p=-r$.  In the remaining diagrams, the on-shell legs are
distributed into two 
subsets $\Oper(r)=\Oper_1(r_1)\Oper_2(r_2)$ with momenta $r_1+r_2=r$.
One subset $\Oper_1$ interacts with the leg $\varphi^+_{i'}$, which
receives momentum $p+r_2=-r_1$. The other subset $\Oper_2$ interacts
with the ghost line. Therefore, in LO the last term on the l.h.s.\ of
\refeq{PSWISCABOS} yields
\beqar 
\lefteqn{-\ri e\sum_{V^b,\varphi_{i'}}
\int \ddl
\langle\bar{u}^{a}(q)u^{b}(l)\varphi^+_{i'}(p-q-l)\Oper(r)\rangle
I^{V^b}_{\varphi_{i'}\varphi_i}
}\quad\nl
&=&
-\ri e\sum_{\varphi_{i'}}
\langle\bar{u}^{a}(q)u^{a}(-q)\rangle
\langle\varphi^+_{i'}(p)\Oper(r)\rangle
I^{V^a}_{\varphi_{i'}\varphi_i}
\nl 
&&{}-\ri e\sum_{V^b,\varphi_{i'}}
\sum_{\Oper_1\neq\Oper}
\langle\bar{u}^{a}(q)u^{b}(-q-r_2)\Oper_2(r_2)\rangle
\langle\varphi^+_{i'}(p+r_2)\Oper_1(r_1)\rangle
I^{V^b}_{\varphi_{i'}\varphi_i}, \eeqar and if we split off the
momentum-conservation $\de$-functions, \refeq{PSWISCABOS} becomes
\beqar\label{SCABRSWI}
\lefteqn{\frac{\ri}{\xi_a} q^\mu G^{\bar{V}^a {\varphi^+_i} \underline{\Oper}}_{\mu}(q,p-q,r)
-\ri e v \sum_{\Phi_j}I^{V^a}_{H\Phi_j}
G^{{\Phi}_j {\varphi^+_i} \underline{\Oper}}(q,p-q,r)}\quad\nl
\nl&&{}
+\sum_{V^b} \linbrs^{V^b}_{\varphi^+_i}
G^{\bar{u}^a {u}^b \underline{\Oper}}(q,p-q,r)
-\ri e \sum_{\varphi_{i'}}
G^{\bar{u}^a u^a}(q)G^{{\varphi^+_{i'}} \underline{\Oper}}(p,r)
I^{V^a}_{\varphi_{i'}\varphi_i} 
\nl&=&
\ri e\sum_{V^b,\varphi_{i'}} \sum_{\Oper_1\neq \Oper}
G^{{\varphi^+_{i'}} \underline{\Oper}_1}(p+r_2,r_1)
I^{V^b}_{\varphi_{i'}\varphi_i}
G^{\bar{u}^a {u}^b\underline{\Oper}_2}(q,-q-r_2,r_2)
. 
\eeqar
Recall that we are interested in the on-shell and ``massless'' limit
$p^2\ll s$ of the above equation. Therefore, we have to take special
care of all terms that are enhanced in this limit, like internal
propagators carrying momentum $p$.  Since internal lines with small
invariants do not occur on the r.h.s.\ of \refeq{SCABRSWI}, we now
concentrate on the l.h.s.  Using \refeq{subtractGF}, the first term
can be written as
\beqar
G_\mu^{\bar{V}^a{\varphi^+_i} \underline{\Oper}}(q,p-q,r) 
&=&G_\mu^{[\bar{V}^a{\varphi^+_{i}}] \underline{\Oper}}(q,p-q,r) 
\nl&&{}
+\sum_{\varphi_{i'}}G_\mu^{\bar{V}^a {\varphi^+_i} \varphi_{i'}}(q,p-q,-p)G^{\underline{\varphi}_{i'} \underline{\Oper}}(p,r)
\nl&&{}
+ \sum_{\tilde{\varphi}_j}G^{\bar{V}^a {\varphi^+_i} \tilde{\varphi}_j}_{\mu}(q,p-q,-p)G^{ \underline{\tilde{\varphi}_j} \underline{\Oper}}(p,r)
,
\eeqar
where for scalar $\varphi^+_i$ the sums run over scalar $\varphi_{i'}$
and vector $\tilde{\varphi}_j$ and vice versa if $\varphi^+_i$ is a
vector.
In this way the enhanced internal propagators with momentum $p$ are
isolated in the terms $G^{\bar{V}^a \varphi^+_i
  \varphi_{i'}}_{\mu}(q,p-q,-p)$ and $G^{\bar{V}^a \varphi^+_i
  \tilde{\varphi}_j}_{\mu}(q,p-q,-p)$, whereas the subtracted Green
functions $G_\mu^{[\bar{V}^a{\varphi}_{i}^+] \underline{\Oper}}$ contain
no enhancement by definition. A similar decomposition is used for the
second and third term on the l.h.s.\ of \refeq{SCABRSWI}, whereas the
enhanced propagator contained in the last term is isolated by writing
\beq
G^{{\varphi^+_{i'}} \underline{\Oper}}(p,r)=
G^{{\varphi^+_{i'}}{\varphi}_{i'}}(p)
G^{\underline{\varphi}_{i'} \underline{\Oper}}(p,r).
\eeq
In this way, the l.h.s.\ of \refeq{SCABRSWI} can be written as
\beqar\label{SCABRSWIb}
\lefteqn{
\frac{\ri}{\xi_a} q^\mu G^{[\bar{V}^a {\varphi^+_i}] \underline{\Oper}}_{\mu}(q,p-q,r)
-\ri e v \sum_{\Phi_j}I^{V^a}_{H\Phi_j}
G^{[{\Phi}_j {\varphi^+_i}] \underline{\Oper}}(q,p-q,r)
}\quad
\nl&&
{}+\sum_{V^b} \linbrs^{V^b}_{\varphi^+_i}
G^{[\bar{u}^a {u}^b] \underline{\Oper}}(q,p-q,r)
+\sum_{\varphi_{i'}}S^{\bar{V}^a}_{\varphi^+_{i}\varphi_{i'}}
G^{\underline{\varphi}_{i'} \underline{\Oper}}(p,r)
+\sum_{\tilde{\varphi}_j}M^{\bar{V}^a}_{\varphi^+_{i}\tilde{\varphi}_j}
G^{\underline{\tilde{\varphi}}_j \underline{\Oper}}(p,r),\nln
\eeqar
where all enhanced terms are in the self-energy-like ($\varphi\varphi$) 
contributions
\beqar\label{SCASterm}
S^{\bar{V}^a}_{\varphi^+_{i}\varphi_{i'}}&=&
\frac{\ri}{\xi_a} q^\mu G^{\bar{V}^a {\varphi^+_i} \varphi_{i'}}_{\mu}(q,p-q,-p)
-\ri e v \sum_{\Phi_k}I^{V^a}_{H\Phi_k}
G^{{\Phi}_k {\varphi}^+_i \varphi_{i'}}(q,p-q,-p)
\nl&&
{}+ \sum_{V^b} \linbrs^{V^b}_{\varphi^+_i}
G^{\bar{u}^a {u}^b \varphi_{i'}}(q,p-q,-p)
-\ri e G^{\bar{u}^a u^a}(q)
G^{{\varphi^+_{i'}}{\varphi}_{i'}}(p)
I^{V^a}_{\varphi_{i'}\varphi_i}
\eeqar
and in the mixing-energy-like ($\varphi\tilde\varphi$)
 contributions
\beqar\label{SCAMterm}
M^{\bar{V}^a}_{\varphi^+_{i}\tilde{\varphi}_j}
&=&\frac{\ri}{\xi_a} q^\mu G^{\bar{V}^a {\varphi^+_i}\tilde{\varphi}_j}_{\mu}(q,p-q,-p)
-\ri e v \sum_{\Phi_k}I^{V^a}_{H\Phi_k} G^{{\Phi}_k {\varphi^+_i} \tilde{\varphi}_j}(q,p-q,-p)
\nl&&
{}+\sum_{V^b} \linbrs^{V^b}_{\varphi^+_i}G^{\bar{u}^a {u}^b \tilde{\varphi}_j}(q,p-q,-p).\hspace{4cm}
\eeqar
Note that here the terms originating from $\L_v$, \ie terms
proportional to the vev, are enhanced by the internal
$\tilde\varphi_j$ propagators and represent leading contributions to
\refeq{SCABRSWIb}.  In order to simplify \refeq{SCASterm} and
\refeq{SCAMterm}, and to check whether contributions proportional to
the vev survive, we have to derive two further Ward identities.

\begin{itemize}
\item For the self-energy-like
 contributions \refeq{SCASterm}  we exploit the BRS invariance of the
 Green function $\langle\bar{u}^{a}(x)\varphi^+_{i}(y)\varphi_{i'}(z)\rangle$:
\beq
\langle[\brs \bar{u}^{a}(x)]\varphi^+_{i}(y)\varphi_{i'}(z)\rangle
-\langle\bar{u}^{a}(x)[\brs \varphi^+_{i}(y)]\varphi_{i'}(z)\rangle
=\langle\bar{u}^{a}(x)\varphi^+_{i}(y)[\brs \varphi_{i'}(z)]\rangle.
\eeq 
Using the BRS variations \refeq{antighostbrstra},
\refeq{ccphysbrstra}, and \refeq{physbrstra}, we have
\beqar
\lefteqn{\frac{1}{\xi_a}\partial^\mu_{x}\langle\bar{V}^a_\mu(x)\varphi^+_{i}(y)\varphi_{i'}(z)\rangle
-\ri e v \sum_{\Phi_j}I^{V^a}_{H\Phi_j}\langle\Phi_j(x) \varphi^+_{i}(y)\varphi_{i'}(z)\rangle}\quad
\nl&&{}
+\sum_{V^b}\linbrs^{V^b}_{\varphi^+_i}
\langle\bar{u}^{a}(x)u^{b}(y)\varphi_{i'}(z)\rangle
-\ri e\sum_{V^b,\varphi_k}
\langle\bar{u}^{a}(x)u^{b}(y)\varphi^+_{k}(y)\varphi_{i'}(z)\rangle I^{V^b}_{\varphi_{k}\varphi_i}
\nl
&=&
-\sum_{V^b}\linbrs^{V^b}_{\varphi_{i'}}
\langle\bar{u}^{a}(x)\varphi^+_{i}(y)u^{b}(z)\rangle
-\ri e\sum_{V^b,\varphi_k}
I^{V^b}_{\varphi_{i'}\varphi_{k}}\langle\bar{u}^{a}(x)\varphi^+_{i}(y)u^{b}(z)\varphi_{k}(z)\rangle.\nln
\eeqar
In LO, the terms involving four fields reduce to  products of pairs of
propagators. After Fourier transformation we obtain  
\beqar\label{vertexWI}
\lefteqn{\frac{\ri}{\xi_a} q^\mu\langle\bar{V}^a_\mu(q)\varphi^+_{i}(p-q)\varphi_{i'}(-p)\rangle
-\ri e v \sum_{\Phi_j}I^{V^a}_{H\Phi_j}\langle\Phi_j(q) \varphi^+_{i}(p-q)\varphi_{i'}(-p)\rangle}\quad
\nl&&{}
+\sum_{V^b}\linbrs^{V^b}_{\varphi^+_i}
\langle\bar{u}^{a}(q)u^{b}(p-q)\varphi_{i'}(-p)\rangle
-\ri e  \langle\bar{u}^{a}(q)u^{a}(-q)\rangle
\langle\varphi^+_{i'}(p)\varphi_{i'}(-p)\rangle I^{V^a}_{\varphi_{i'}\varphi_i}
\nl&=&
-\sum_{V^b}\linbrs^{V^b}_{\varphi_{i'}}
\langle\bar{u}^{a}(q)\varphi^+_{i}(p-q)u^{b}(-p)\rangle
\nl&&{}
-\ri e \langle\bar{u}^{a}(q)u^{a}(-q)\rangle \langle\varphi^+_{i}(p-q) \varphi_{i}(-p+q)\rangle I^{V^a}_{\varphi_{i'}\varphi_i}
,
\eeqar
and we easily see that 
\beq  \label{CWISCABOSamp}
S^{\bar{V}^a}_{\varphi^+_{i}\varphi_{i'}}
=
-\sum_{V^b} \linbrs^{V^b}_{\varphi_{i'}}G^{\bar{u}^a {\varphi}^+_{i} u^b}(q,p-q,-p)
-\ri e 
G^{\bar{u}^a u^a}(q)G^{\varphi^+_i\varphi_i}(p-q)
I^{V^a}_{\varphi_{i'}\varphi_i}.
\eeq

\item For the mixing-energy-like contributions \refeq{SCAMterm} we use
  the BRS invariance of the Green function
  $\langle\bar{u}^{a}(x)\varphi^+_{i}(y)\tilde{\varphi}_j (z)\rangle$.
  The resulting WI is obtained from \refeq{vertexWI} by substituting
  $\varphi_{i'}\rightarrow \tilde{\varphi}_j$ and by neglecting the
  mixing propagators $\langle\varphi^+_{i}(p)\tilde{\varphi}_j
  (-p)\rangle$ which vanish in LO and reads
\beqar\label{brokenWIs}
\lefteqn{\frac{\ri}{\xi_a} q^\mu\langle\bar{V}^a_\mu(q)\varphi^+_{i}(p-q)\tilde{\varphi}_{j}(-p)\rangle
-\ri e v \sum_{\Phi_k}I^{V^a}_{H\Phi_k}\langle\Phi_k(q) \varphi^+_{i}(p-q)\tilde{\varphi}_{j}(-p)\rangle}\quad
\\&&{}
+\sum_{V^b}\linbrs^{V^b}_{\varphi^+_i}
\langle\bar{u}^{a}(q)u^{b}(p-q)\tilde{\varphi}_{j}(-p)\rangle
= -\sum_{V^b}\linbrs^{V^b}_{\tilde{\varphi}_{j}}
\langle\bar{u}^{a}(q)\varphi^+_{i}(p-q)u^{b}(-p)\rangle.\nonumber
\eeqar
This relation involves the $VV\Phi$ couplings as well as  other terms originating from  $\L_{v}$, and leads to
\beq  \label{CWISCABOS2}
M^{\bar{V}^a}_{\varphi^+_i\tilde{\varphi}_j}
=-\sum_{V^b} \linbrs^{V^b}_{\tilde{\varphi}_j} G^{\bar{u}^a {\varphi}^+_{i} u^b}(q,p-q,-p).
\eeq
\end{itemize}
Both \refeq{CWISCABOSamp} and \refeq{CWISCABOS2} contain the ghost
vertex function $G^{\bar{u}^a {\varphi}^+_{i} u^b}$, but when we
combine them in \refeq{SCABRSWIb} these ghost contributions cancel
owing to the LO identity that 
relates external would-be Goldstone bosons and gauge bosons,
\beqar 
\lefteqn{\left[\sum_{V^d} 
G_\mu^{\underline{V}^d \underline{\Oper}}(p,r)\linbrs^{V^b}_{V^d_\mu}
+ \sum_{\Phi_j}G^{\underline{\Phi}_{j} \underline{\Oper}}(p,r)\linbrs^{V^b}_{\Phi_j} \right]G^{\bar{u}^a {\varphi}^+_{i} u^b}(q,p-q,-p)}\quad
\nl&=& 
\left[-\ri p^\mu G_\mu^{\underline{V}^{b} \underline{\Oper}}(p,r)
+\ri e v \sum_{\Phi_{j}} I^{V^b}_{\Phi_{j}H} G^{\underline{\Phi}_{j}
  \underline{\Oper}}(p,r)\right]G^{\bar{u}^a {\varphi}^+_{i} u^b}(q,p-q,-p)
\nl&=&0.  
\eeqar
Thus, all mixing terms cancel and the complete identity
\refeq{SCABRSWI} becomes
\beqar \label{CWISCABOSa0}
\lefteqn{
\frac{\ri}{\xi_a} q^\mu G^{[\bar{V}^a {\varphi^+_i}]
  \underline{\Oper}}_{\mu}(q,p-q,r) -\ri e v
\sum_{\Phi_j}I^{V^a}_{H\Phi_j} G^{[{\Phi}_j {\varphi^+_i}]
  \underline{\Oper}}(q,p-q,r)}\quad
 \nl&&{} 
+\sum_{V^b}
\linbrs^{V^b}_{\varphi^+_i} G^{[\bar{u}^a {u}^b]
  \underline{\Oper}}(q,p-q,r) - \ri e G^{\bar{u}^a
  u^a}(q)G^{\varphi^+_i\varphi_i}(p-q)
\sum_{\varphi_{i'}}G^{\underline{\varphi}_{i'} \underline{\Oper}}(p,r)
I^{V^a}_{\varphi_{i'}\varphi_i} \nl
&=& \ri e\sum_{V^b,\varphi_{i'}}
\sum_{\Oper_1\neq \Oper} G^{{\varphi}^+_{i'}
  \underline{\Oper}_1}(p+r_2,r_1) I^{V^b}_{\varphi_{i'}\varphi_i}
G^{\bar{u}^a {u}^b\underline{\Oper}_2}(q,-q-r_2,r_2) .  
\eeqar 
Now we can truncate the two remaining external legs.  To this end we
observe that (see \refapp{app:GFs}) the longitudinal part of the LO
gauge-boson propagator $G_\rL^{V^a\bar{V}^a}(q)$, the LO ghost
propagator, and the LO propagator of the associated would-be Goldstone
boson $\Phi_j$ are related by
\beq
\frac{1}{\xi_a}
G_\rL^{V^a\bar{V}^a}(q)=G^{\bar{u}^au^a}(q)=-G^{\Phi^+_j\Phi_j}(q).
\eeq   
Using this relation, the leg with momentum $q$ is easily truncated by
multiplying the above identity with the longitudinal part of the
inverse gauge-boson propagator $-\ri\xi_a
\Gamma^{V^a\bar{V}^a}_\rL(q)$.  The leg with momentum $p-q$ is
truncated by multiplying \refeq{CWISCABOSa0} by the inverse
(scalar-boson or gauge-boson) propagator $-\ri
\Gamma^{\varphi_i\varphi_i^+}(p-q)$, and by using
\beq
\linbrs^{V^b}_{\varphi^+_i}G^{u^b\bar{u}^b}(p-q)=
c_{\varphi_i}G^{\varphi^+_i\varphi_i}(p-q)\linbrs^{V^b}_{\varphi^+_i}.
\eeq
with $c_{\Phi_i}=1$ and  $c_{V^b}=-1/\xi_b$.
The truncated identity reads
\beqar \label{CWISCABOSa}
\lefteqn{
\ri q^\mu G^{[\underline{V}^a \underline{\varphi}_i] \underline{\Oper}}_{\mu}(q,p-q,r)
+\ri e v \sum_{\Phi_j}I^{V^a}_{H\Phi_j}
G^{[\underline{\Phi}_j^+ \underline{\varphi}_i] \underline{\Oper}}(q,p-q,r)
}\quad
\nl&&{}
+\sum_{V^b} c_{\varphi_i} \linbrs^{V^b}_{\varphi^+_i}
G^{[\underline{u}^a \underline{\bar{u}}^b] \underline{\Oper}}(q,p-q,r)
-
\ri e \sum_{\varphi_{i'}}
G^{\underline{\varphi}_{i'} \underline{\Oper}}(p,r)
I^{V^a}_{\varphi_{i'}\varphi_i} 
\nl&=&
\ri e\sum_{V^b,\varphi_{i'}} \sum_{\Oper_1\neq \Oper}
\left[-\ri \Gamma^{\varphi_i\varphi_i^+}(p-q)
G^{{\varphi}^+_{i'} \underline{\Oper}_1}(-r_1,r_1) \right]
I^{V^b}_{\varphi_{i'}\varphi_i}
G^{\underline{u}^a{u}^b\underline{\Oper}_2}(q,-q-r_2,r_2).
\nln
\eeqar
Now we contract with the wave function $v_{\varphi}(p)$ of an on-shell
external state with mass $\sqrt{p^2}$.
For scalar bosons the wave function is trivial ($v_\Phi(p)=1$),
whereas for external gauge bosons we consider transverse polarizations
$v_{V_\nu}(p)=\varepsilon_\rT^\nu(p)$.  Finally, when we take the
collinear limit
$q^\mu \rightarrow xp^\mu$ and assume 
\beq
M^2\sim\max(p^2,M_{\varphi_i}^2,M_{V^a}^2)\ll s,
\eeq
various terms in \refeq{CWISCABOSa} are
mass-suppressed.  The r.h.s.\ is mass-suppressed owing to
\beq\label{rhsmsupp}
\lim_{q^\mu \rightarrow xp^\mu}
v_{\varphi}(p)\Gamma^{\varphi_i\varphi^+_i}(p-q)
G^{\varphi^+_{i'}\varphi_{i'}}(-r_1)
\sim\lim_{q^\mu \rightarrow xp^\mu}
\frac{(p-q)^2-M_{\varphi_i}^2}{r_1^2}
=\O\left(\frac{M^2}{s}\right),
\eeq
since $r_1$ is a non-trivial combination of the external momenta, and
like for all invariants we assume that $r_1^2\sim s$, whereas in the
collinear limit $(p-q)^2-M_{\varphi_i}^2\sim M^2$.
  The second term on the the l.h.s.\ 
of \refeq{CWISCABOSa} is proportional to the vev and therefore
mass-suppressed, and for the third term we have
\beq
\lim_{q^\mu \rightarrow xp^\mu} v_{\varphi}(p) \linbrs^{V^b}_{\varphi^+_i}=\O(M).
\eeq
For gauge bosons this is due to the transversality of the polarization vector
\beq
\lim_{q^\mu \rightarrow xp^\mu}\, (p-q)_\nu\varepsilon_\rT^\nu(p)=
0,
\eeq
whereas for scalar bosons $\linbrs^{V^b}_{\Phi^+_i}$ is explicitly
proportional to the vev.  The remaining leading terms give the result
\beqar \label{SCACWIres}
\lefteqn{\lim_{q^\mu \rightarrow xp^\mu}
q^\mu v_{\varphi}(p)
 G^{[\underline{V}^a \underline{\varphi}_i]
   \underline{\Oper}}_{\mu}(q,p-q,r)}\quad
\nl&=&
e \sum_{\varphi_{i'}}
v_{\varphi}(p)G^{\underline{\varphi}_{i'} \underline{\Oper}}(p,r)
I^{V^a}_{\varphi_{i'}\varphi_i} 
+\O\left(\frac{M^2}{s} \M^{\varphi_{i}\Oper}\right),
\nl&=&
e \sum_{\varphi_{i'}}
\M^{\varphi_{i'}\Oper}(p,r)
I^{V^a}_{\varphi_{i'}\varphi_i} 
+\O\left(\frac{M^2}{s} \M^{\varphi_{i}\Oper}\right),
\eeqar
which is the identity represented in \refeq{SCACWIdiag} in diagrammatic form.
Note that in general 
the mass of the wave
function $v_{\varphi}(p)$ need not be equal to the masses of the fields
$\varphi_i$ or $\varphi_{i'}$.

\subsection{Fermions}\label{se:FERCWI}
The derivation of the collinear Ward identity for external fermions
$\varphi_i=\Psi^\kappa_{j,\si}$  is completely analogous to those
presented in the previous section. 
In fact, it is  much simpler since no mixing contributions
($\tilde{\varphi}$) have to be considered. The effect of the
quark-mixing matrix can be absorbed in the generalized generators 
\beq
I^{V^a}_{\varphi_i\varphi_{i'}}=U^{V^a}_{jj'}I^{V^a}_{\si\si'}.
\eeq
The final result reads
\beqar\label{FERCWIres}
\lim_{q^\mu \rightarrow xp^\mu} 
q^\mu G^{[\underline{V}^a\underline{\Psi}_{j,\si}^\kappa] 
\underline{\Oper}}_{\mu}(q,p-q,r)u(p)&=& + e \sum_{\si',j'}
\M^{f_{j',\si'}^\kappa \Oper}(p,r)
I^{V^a}_{\si'\si}U^{V^a}_{j'j}
+\O\left(\frac{M}{\sqrt{s}} \M^{f_{j,\si}^\kappa\Oper}\right),\nln
\eeqar
for fermions, and
\beqar\label{ANTIFERCWIres}
\lim_{q^\mu \rightarrow xp^\mu}
q^\mu \bar{v}(p)
G^{[\underline{V}^a\underline{\bar{\Psi}}_{j,\si}^\kappa] 
\underline{\Oper}}_{\mu}(q,p-q,r)&=& -e \sum_{\si',j'}I^{V^a}_{\si\si'}U^{V^a}_{jj'}
\M^{\bar{f}_{j',\si'}^\kappa \Oper}(p,r)
+\O\left(\frac{M}{\sqrt{s}} \M^{\bar{f}_{j,\si}^\kappa\Oper}\right),\nln
\eeqar
for antifermions,
where
\beq
M^2\sim\max{(p^2, m_{f_{j,\si}}^2, M_{V^a}^2)}.
\eeq
In the derivation we used
\beq
\lim_{q^\mu \rightarrow xp^\mu} \bar{v}(p)
\Gamma^{\bar{\Psi}\Psi}_{j,\si}(p-q) G^{\Psi\bar{\Psi}}_{j',\si'}(-r_1)
\propto \frac{M\bar{v}(p)\rs_1}{r_1^2} =\O\left(\frac{M}{\sqrt{s}}\right),
\eeq
which is the analogue of \refeq{rhsmsupp}, and a similar equation for
fermion spinors $u(p)$.

\section{Conclusions}

For energies at and beyond $1\TeV$, the electroweak corrections are
dominated by double and single logarithms involving the ratio of the
typical energy of the considered process and the electroweak scale.
For processes that are not mass-suppressed, the one-loop logarithmic
corrections are universal, \ie in contrast to the non-logarithmic
corrections they can be calculated in a process-independent way.  The
corresponding results have already been published in
\citere{Denner:2001jv}.

Here we have presented the derivation of the virtual collinear
logarithms at the one-loop level in the electroweak Standard Model for
processes that are not mass-suppressed. Using the BRS invariance of
the electroweak Standard Model, we have proved the factorization of
these logarithms in the 't~Hooft--Feynman gauge.  The proof has been
performed in the spontaneously broken phase in terms of the physical
fields and parameters. The mixings between the various fields and all
relevant terms proportional to the vacuum expectation value have been
taken into account.  We find that all terms proportional to the vacuum
expectation value cancel and the results are equivalent to those
obtained in the symmetric phase with the longitudinal modes of the
gauge bosons replaced by the would-be Goldstone bosons as physical
particles. Thus, this equivalence, which has been assumed in the
literature, has been proven at the one-loop level using the
Goldstone-boson equivalence theorem and the corresponding corrections.
It will be interesting to investigate to what extent this equivalence
is valid at higher orders.

While we have derived the collinear Ward identities and the collinear
logarithms within the Electroweak Standard Model our method can be
generalized to arbitrary spontaneously broken gauge theories
including, in particular, supersymmetric extensions of the Standard
Model.

\appendix

\section{Collinear singularity}\label{app:masssing} 
In this appendix we discuss mass singularities originating from
integrals of the type
\beqar \label{masssingloop3}
I=-\ri{(4\pi)^2\mu^{4-D}}
\int\ddq \frac{N(q)}{(q^2-M_0^2+\ri\varepsilon)[(p-q)^2-M_1^2+\ri\varepsilon]}.
\eeqar
We restrict ourselves to purely collinear singularities that are
exclusively related to an external relativistic momentum $p^\mu$ that
has a small square,
\ie $p^2\ll(p^0)^2\sim s$.  Singularities originating from other
propagators in $N(q)$ are not considered.
In particular, we assume that $N(q)$ is either 
not singular in the
soft limit $q^\mu\rightarrow 0$ or that the soft singularities are
subtracted.

Our goal is to fix a precise prescription for 
extracting the part of the function $N(q)$ 
that enters the mass-singular part of
\refeq{masssingloop3}. To this end, we introduce a Sudakov
parametrization  \cite{SUD}
for the loop momentum  
\newcommand{\light}{l}
\beq\label{Sudparam}
q^\mu=xp^\mu+y \light^\mu +q^\mu_\rT,
\eeq
where $p^\mu$ and the light-like four vector $\light^\mu$,
\beq
p^\mu=(p^0,\vec{p}),\qquad
 \light^\mu=(p^0,-p^0\vec{p}/|\vec{p}|),
\eeq
describe the component collinear to the external momentum, whereas the
space-like vector $q^\mu_\rT$ with   
\beq
p_\mu q^\mu_\rT =\l_\mu q^\mu_\rT=0,\qquad q_\rT^2=-|\vec{q}_\rT|^2
\eeq
represents the perpendicular component. In this parametrization we get 
\beq \label{masssingloop4}
I=
-4\ri(p\light) \mu^{4-D}
\int\mathrm{d}x\int\mathrm{d}y \int\ddqt \frac{N(q)}{(q^2-M_0^2+\ri\varepsilon)[(p-q)^2-M_1^2+\ri\varepsilon]}.
\eeq
The denominators of the propagators read
\beqar
q^2-M_0^2+\ri\varepsilon&=&x^2p^2+2xy(p\light)-|\vec{q}_\rT|^2-M_0^2+\ri\varepsilon,
\nl
(p-q)^2-M_1^2+\ri\varepsilon&=&(1-x)^2p^2+2(x-1)y(p\light)-|\vec{q}_\rT|^2-M_1^2+\ri\varepsilon,
\eeqar
and are linear in the the variable $y$ owing to $\l^2=0$. 
For $x\ne0,1$, the integral $I$ can be written as 
\beq \label{masssingloop5}
I=
-\ri\frac{\mu^{4-D}}{(p\light)}
\int\frac{\mathrm{d}x}{x(x-1)} \int\ddqt \int\mathrm{d}y\, \frac{N(x,y,q_\rT)}{(y-y_0)(y-y_1)}
\eeq
with single poles at
\beqar\label{poles}
y_0&=&\frac{|\vec{q}_\rT|^2-x^2p^2+M_0^2-\ri\varepsilon}{2x(p\light)},\qquad x\neq 0,
\nl
y_1&=&\frac{|\vec{q}_\rT|^2-(1-x)^2p^2+M_1^2-\ri\varepsilon}{2(x-1)(p\light)},\qquad x\neq 1.
\eeqar
The  $y$~integral is non-zero only when the poles lie in opposite
complex half-planes, \ie for $0<x<1$. Then, it can be
performed by closing the contour around one of the two poles. This yields  
\beqar \label{masssingloop6}
I&=&-\frac{2\pi\mu^{4-D}}{(p\light)}\int_0^1\frac{\mathrm{d}x}{x(x-1)} \int\ddqt  \frac{ N(x,y_i,q_\rT)}{y_0-y_1}
\nl&=&
4\pi\mu^{4-D}\int_0^1\mathrm{d}x \int\ddqt \frac{N(x,y_i,q_\rT)}{|\vec{q}_\rT|^2+\Delta(x)},
\eeqar
where in the vicinity of $x=1,0$  the contour has to be closed around the pole at $y_i=y_0,y_1$, respectively. 

The transverse momentum integral exhibits a
logarithmic singularity in the collinear region $|q_\rT|\rightarrow
0$, where
the squares of the momenta $p$ and $p-q$ 
are small compared to 
the energy squared $p^2,(p-q)^2\ll (p\light)\sim 2 p_0^2$. 
The singularity is regulated by the mass terms in 
\beq\label{Delta}
\Delta(x)=(1-x)M_0^2+xM_1^2-x(1-x)p^2.
\eeq

In leading approximation, we restrict ourselves to logarithmic
mass-singular contributions in \refeq{masssingloop6}.  Terms 
of order $|\vec{q}_\rT|^2$, $p^2$, $M_0$ or $M_1$ 
are neglected in $N(q)$.
Since the relevant pole, $y_0$ or
$y_1$, is of order $|\vec{q}_\rT|^2/(p\light)$,
also contributions proportional to $y$ can be discarded. We therefore arrive
at the following simple 
recipe for $N(q)$ in the collinear limit:
\beqar\label{collappdef}
&(1)\,&\mbox{Substitute $N(x,y,q_\rT)\rightarrow N(x,0,0)$, \ie  replace  $q^\mu \rightarrow xp^\mu$.}\hspace{3cm}
\nl&(2)\,&
\mbox{Neglect all mass-suppressed contributions.}
\eeqar
Then, performing  the $q_\rT$ integration in $D-2=2-2\varepsilon$
dimensions and expanding in $\varepsilon$, we obtain the leading
contribution 
\beqar \label{masssingloop7}
I&=&\Gamma(\varepsilon)\int_0^1\mathrm{d}x \left(\frac{4\pi\mu^2}{\De(x)}\right)^\varepsilon N(x,0,0)
\nl&=&
\frac{1}{\varepsilon}+\int_0^1\mathrm{d}x\, \log{\left(\frac{\mu^2}{\De(x)}\right)} N(x,0,0)-\gamma+\log{4\pi}+\O(\varepsilon).
\eeqar
Finally, omitting the UV singularity, which cancels in observables,
neglecting constant terms, and performing the integral, we obtain
\beq \label{masssingloop8}
I\LA \log{\left(\frac{\mu^2}{M^2}\right)}\int_0^1\mathrm{d}x\,  N(x,0,0),
\eeq
 in logarithmic approximation (LA).
The scale in the logarithm is of the order of 
the largest mass in
\refeq{Delta}, 
\beq
M^2\sim\max{(p^2,M_0^2,M_1^2)}.
\eeq

\section{BRS transformations} \label{BRStra}
In this appendix we summarize our conventions for 
the gauge-fixing terms and the BRS symmetry of the electroweak
Standard Model. We follow
\citere{DennBohmJos} but introduce a generic notation.
\subsection{Gauge symmetry}
The classical Lagrangian of the electroweak Standard Model is
invariant with respect to gauge transformations of the physical fields
(and would-be Goldstone bosons) $\varphi_i$, which can generically be
written as
\beq \label{physgaugetra}
\de \varphi_i(x)= \sum_{V^b=A,Z,W^\pm}\left[\linbrs^{V^b}_{\varphi_{i}}\de\theta^{V^b}(x)
+\ri e  \sum_{\varphi_{i'}}
I^{V^b}_{\varphi_i\varphi_{i'}}\de\theta^{V^b}(x) \varphi_{i'}(x)\right].
\eeq
The linear operator $\linbrs^{V^b}_{\varphi_{i}}$ represents the
transformation of free fields, and the non-linear
term contains the
$\ewgroup$ generators $I^{V^b}_{\varphi_i\varphi_{i'}}$ in the
representation of the fields $\varphi_{i}$ \cite{Denner:2001jv}. 
For scalar bosons,
the linear term in \refeq{physgaugetra} is determined by the
contribution of the vev
\beq
\vev_i=v \, \de_{H\Phi_{i}}
\eeq
and reads
\beq
\linbrs^{V^b}_{\Phi_{i}}\de\theta^{V^b}(x)=\ri ev
I^{V^b}_{\Phi_iH}\de\theta^{V^b}(x). 
\eeq
For gauge bosons, $\varphi_i=V^b=A,Z,W^\pm$, we have
\beq
\linbrs^{V^b}_{V^c_\mu}\de\theta^{V^b}(x)=\de_{V^bV^c}\partial_\mu \de\theta^{V^b}(x),
\eeq
which in momentum space leads to the simple relation
\beq
\linbrs^{V^b}_{V^c_\mu}\de\theta^{V^b}(p) = \ri p_\mu\de_{V^bV^c}\de\theta^{V^b}(p) .
\eeq
For fermions, $\varphi_i=\Psi^\kappa_{j,\si}$,
\beq
\linbrs^{V^b}_{\Psi_{j,\si}}\de\theta^{V^b}(x)=0,
\eeq
and the gauge transformation of the physical 
fields is determined by
\beq
I^{V^a}_{\Psi_{j,\si}\Psi_{j',\si,}}=U^{V^a}_{jj'}I^{V^a}_{\si\si'},
\eeq
where the generators $I^{V^a}_{\si\si'}$ depend on the representation
of $\Psi^\kappa_{j,\si}$ and, 
in particular,
on the chirality
$\kappa=\rR,\rL$. 
The unitary mixing matrix $U^{V^a}_{jj'}$ is trivial
$(U^{V^a}_{jj'}=\de_{jj'})$ everywhere except for the left-handed quark
representation, where it has the non-trivial components
\beq\label{mixingmatrix}
U^{W^+}_{jj'}=\ckm_{jj'},\qquad U^{W^-}_{jj'}=\ckm^+_{jj'}=\ckm^*_{j'j},
\eeq
corresponding to the quark-mixing 
matrix $\ckm_{jj'}$.

\subsection{Gauge fixing and BRS invariance}
The quantized
electroweak Lagrangian includes the gauge-fixing term
\beq
\mathcal{L}_{\mathrm{fix}}=
-\sum_{V^a=A,Z,W^\pm}\frac{1}{2\xi_a}C^{{V}^a}C^{\bar{V}^a},
\eeq
with the gauge parameters  $\xi_A$, $\xi_Z$, $\xi_+=\xi_-$, 
and the corresponding ghost terms.
The charge-conjugate of $V$ is denoted $\bar V$.
A general 't Hooft gauge fixing is given by
\beq  \label{Gfix}
C^{\bar{V}^a}\{V,\Phi,x\}=
\partial^\mu \bar{V}^a_\mu(x)-
\ri e v \xi_{{a}}\sum_{\Phi_{i}=H,\chi,\phi^\pm}I^{V^a}_{H\Phi_i}\Phi_i(x).
\eeq
Note that the matrix elements
$I^{V^a}_{H\Phi_i}$ relate the gauge fields $V^a$ to the associated
would-be
Goldstone-boson fields $\Phi_i$. In fact, the single components of
\refeq{Gfix} read
\beqar
C^A(x)&=&\partial^\mu A_\mu(x),\qquad
C^Z(x)=\partial^\mu Z_\mu(x)-\xi_Z\MZ\chi(x),\nl
C^{\pm}(x)&=&\partial^\mu W^\pm_\mu(x)\mp\ri \xi_{\pm}\MW \phi^\pm.
\eeqar
In the 't~Hooft gauge 
the contributions of the would-be
Goldstone bosons to the gauge-fixing terms cancel the LO
mixing between gauge bosons and would-be Goldstone bosons.

The  gauge-fixing terms and the ghost terms break the 
gauge invariance of the classical electroweak Lagrangian. However, 
the complete electroweak Lagrangian is invariant 
with respect to BRS transformations of the ghost  and physical fields.

The BRS transformation of the physical fields 
corresponds to a local gauge transformation \refeq{physgaugetra} with
gauge-transformation parameters
$\de\theta^{V^a}(x)=\de\la u^a(x)$ determined by the ghost
fields $u^a(x)$ and the infinitesimal Grassmann parameter $\de\la$. To
be precise, the BRS variation $\brs \varphi_i(x)$ is defined as
left derivative\footnote{The product rule for a Grassmann left derivative is
$
\brs (\varphi_i\varphi_j) =
(\brs \varphi_i) \varphi_j
+(-1)^{n(\varphi_i)}
\varphi_i \brs \varphi_j ,
$
where $n(\varphi_i)$ is given by the ghost plus the fermion number of
the field $\varphi_i$.} with respect to the Grassmann
parameter $\de \la$, \ie 
$\de\varphi_i(x)=\de\la \,\brs \varphi_i(x)$, and reads
\beq \label{physbrstra}
\brs \varphi_i(x)= \sum_{V^b=A,Z,W^\pm}\left[\linbrs^{V^b}_{\varphi_{i}}u^b(x)
+\ri e  \sum_{\varphi_{i'}}
I^{V^b}_{\varphi_i\varphi_{i'}}u^b(x) \varphi_{i'}(x)\right].
\eeq
The BRS variation for charge-conjugate fields is obtained from the
adjoint of \refeq{physbrstra} as 
\beq \label{ccphysbrstra}
\brs \varphi^+_i(x)= \sum_{V^b=A,Z,W^\pm}\left[\linbrs^{V^b}_{\varphi^+_{i}}u^b(x)
-\ri e \sum_{\varphi_{i'}}
u^b(x) \varphi^+_{i'}(x)I^{V^b}_{\varphi_{i'}\varphi_i}\right],
\eeq
where we have used  $\left(I^{V^a}\right)^+=I^{\bar{V}^a}$.

The BRS variation of the ghost fields is given by
\beq  \label{ghostbrstra}
\brs u^b(x)=\frac{\ri e}{2}  \sum_{V^a,V^c=A,Z,W^\pm}
I^{V^a}_{V^bV^c}u^a(x)u^c(x).
\eeq
The BRS variation of the antighost fields is determined by the
gauge-fixing terms, 
\beq  \label{antighostbrstra}
\brs \bar{u}^a(x)=-\frac{1}{\xi_{{a}}}C^{\bar{V}^a}\{V,\Phi,x\}.
\eeq

\section{Conventions for Green functions}

\label{app:GFs} 
Our conventions for Green functions are based on
\citere{DennBohmJos}. In configuration space we use the equivalent
notations 
\beq
G^{\varphi_{i_1}\dots\varphi_{i_n}}(x_1,\dots ,x_n)
=
\langle\varphi_{i_1}(x_1)\dots \varphi_{i_n}(x_n)\rangle.
\eeq
Fourier transformation is defined with incoming momenta, and  the
momentum-conserva\-ti\-on $\de$-function is factorized as  
\beqar
\lefteqn{
(2\pi)^4\de^{(4)}\left(\sum_{k=1}^n p_{k}\right)
G^{\varphi_{i_1}\dots\varphi_{i_n}}(p_1,\dots ,p_n)}\quad
\nl&&=
\int\left(\prod_{k=1}^n 
{\mathrm{d}^4x_k}\right)
\exp{\left(-\ri\sum_{j=1}^n x_jp_j\right)}
G^{\varphi_{i_1}\dots\varphi_{i_n}}(x_1,\dots ,x_n).
\eeqar
Because the field operator $\varphi$ creates antiparticles and
annihilates particles, the fields in the Green
functions 
are associated with outgoing particles (incoming antiparticles).
For propagators we introduce the shorthand notation
\beq
G^{\varphi_{i}\varphi_{j}}(p)=G^{\varphi_{i}\varphi_{j}}(p,-p),
\eeq
For the truncation of the external leg $\varphi_{i_k}$ in momentum
space we adopt the convention  
\beq
G^{\dots\, \varphi_{i_k} \dots}(\dots,p_k,\dots)=
G^{\varphi_{i_k}\varphi^+_{i_k}}(p_k)
G^{\dots \,\underline{\varphi}^+_{i_k} \dots}(\dots,p_k,\dots),
\eeq
where the field argument corresponding to the truncated leg is
underlined and where we have assumed diagonal propagators.  In
truncated Green functions the fields are associated with incoming
particles.

The (diagonal) 
propagators are related to the 2-point vertex functions by
\beq
G^{\varphi_{i}\varphi^+_{i}}(p,-p)\Ga^{\varphi^+_{i}\varphi_{i}}(p,-p)
=\pm\ri
\eeq
with $+$ for scalars and gauge bosons and $-$ for fermions and ghosts.

In the 't~Hooft gauge, the LO propagators are diagonal. They
read
\beq
G^{V^a\bar{V}^b}_{\mu\nu}(p)= 
\left(g_{\mu\nu}-\frac{p_\mu p_\nu}{p^2}\right)G^{V^a\bar{V}^b}_{\rT}(p)
+\frac{p_\mu p_\nu}{p^2}G^{V^a\bar{V}^b}_{\rL}(p),
\eeq
with
\beq
G^{V^a\bar{V}^b}_{\rT}(p)= 
\frac{-\ri\de_{V^aV^b}}{p^2-M_{V^a}^2}
,\qquad
G^{V^a\bar{V}^b}_{\rL}(p)= 
\frac{-\ri\xi_a\de_{V^aV^b}}{p^2-\xi_a M_{V^a}^2},
\eeq
for gauge bosons and
\beq
G^{HH}(p)=\frac{\ri}{p^2-\MH^2},\qquad
G^{\Phi^+_a\Phi_{b}}(p)
= \frac{\ri\de_{V^aV^b}}{p^2-\xi_{a} M_{V_a}^2}
\eeq
for Higgs bosons and would-be 
Goldstone bosons $\Phi_a=\chi,\phi^\pm$
associated to the weak gauge bosons $V^a=Z,W^\pm$.
The propagators for ghost fields are given by
\beq
G^{u^a\bar{u}^b}(-p)=-G^{\bar{u}^bu^a}(p)=\frac{\ri\de_{V^aV^b}}{p^2-\xi_a M_{V^a}^2},
\eeq 
and care must be taken for the sign resulting from the
anticommutativity of the ghost fields.
Similarly, for fermionic fields we have
\beq
G^{\Psi_\alpha\Psibar_\beta}(-p)=-G^{\Psibar_\beta\Psi_\alpha}(p)=\frac{\ri(\ps+m)_{\alpha\beta}}{p^2-m^2},
\eeq 
where $\alpha,\beta$ are Dirac indices.

\section*{Acknowledgements}
This work was supported in part by the Swiss Bundesamt f\"ur Bildung und
Wissenschaft and by the European Union under contract
HPRN-CT-2000-00149. We thank Stefan Dittmaier for a careful reading of
the manuscript.

\newcommand{\vj}[4]{{\sl #1~}{\bf #2~}\ifnum#3<100 (19#3) \else (#3) \fi #4}
 \newcommand{\ej}[3]{{\bf #1~}\ifnum#2<100 (19#2) \else (#2) \fi #3}
 \newcommand{\vjs}[2]{{\sl #1~}{\bf #2}}

 \newcommand{\am}[3]{\vj{Ann.~Math.}{#1}{#2}{#3}}
 \newcommand{\ap}[3]{\vj{Ann.~Phys.}{#1}{#2}{#3}}
 \newcommand{\app}[3]{\vj{Acta~Phys.~Pol.}{#1}{#2}{#3}}
 \newcommand{\cmp}[3]{\vj{Commun. Math. Phys.}{#1}{#2}{#3}}
 \newcommand{\cnpp}[3]{\vj{Comments Nucl. Part. Phys.}{#1}{#2}{#3}}
 \newcommand{\cpc}[3]{\vj{Comp. Phys. Commun.}{#1}{#2}{#3}}
 \newcommand{\epj}[3]{\vj{Eur. Phys. J.}{#1}{#2}{#3}}
 \newcommand{\fp}[3]{\vj{Fortschr. Phys.}{#1}{#2}{#3}}
 \newcommand{\hpa}[3]{\vj{Helv. Phys.~Acta}{#1}{#2}{#3}}
 \newcommand{\ijmp}[3]{\vj{Int. J. Mod. Phys.}{#1}{#2}{#3}}
 \newcommand{\jetp}[3]{\vj{JETP}{#1}{#2}{#3}}
 \newcommand{\jetpl}[3]{\vj{JETP Lett.}{#1}{#2}{#3}}
 \newcommand{\jmp}[3]{\vj{J.~Math. Phys.}{#1}{#2}{#3}}
 \newcommand{\jp}[3]{\vj{J.~Phys.}{#1}{#2}{#3}}
 \newcommand{\lnc}[3]{\vj{Lett. Nuovo Cimento}{#1}{#2}{#3}}
 \newcommand{\mpl}[3]{\vj{Mod. Phys. Lett.}{#1}{#2}{#3}}
 \newcommand{\nc}[3]{\vj{Nuovo Cimento}{#1}{#2}{#3}}
 \newcommand{\nim}[3]{\vj{Nucl. Instr. Meth.}{#1}{#2}{#3}}
 \newcommand{\np}[3]{\vj{Nucl. Phys.}{#1}{#2}{#3}}
 \newcommand{\npbps}[3]{\vj{Nucl. Phys. B (Proc. Suppl.)}{#1}{#2}{#3}}
 \newcommand{\pl}[3]{\vj{Phys. Lett.}{#1}{#2}{#3}}
 \newcommand{\prp}[3]{\vj{Phys. Rep.}{#1}{#2}{#3}}
 \newcommand{\pr}[3]{\vj{Phys.~Rev.}{#1}{#2}{#3}}
 \newcommand{\prl}[3]{\vj{Phys. Rev. Lett.}{#1}{#2}{#3}}                       
 \newcommand{\ptp}[3]{\vj{Prog. Theor. Phys.}{#1}{#2}{#3}}                     
 \newcommand{\rpp}[3]{\vj{Rep. Prog. Phys.}{#1}{#2}{#3}}                       
 \newcommand{\rmp}[3]{\vj{Rev. Mod. Phys.}{#1}{#2}{#3}}                        
 \newcommand{\rnc}[3]{\vj{Revista del Nuovo Cim.}{#1}{#2}{#3}}                 
 \newcommand{\sjnp}[3]{\vj{Sov. J. Nucl. Phys.}{#1}{#2}{#3}}                   
 \newcommand{\sptp}[3]{\vj{Suppl. Prog. Theor. Phys.}{#1}{#2}{#3}}             
 \newcommand{\zp}[3]{\vj{Z. Phys.}{#1}{#2}{#3}}                                
 \renewcommand{\and}{and~}

\end{document}